\documentclass[12pt]{article}
\usepackage{graphicx,floatflt,amssymb,epsf}
\usepackage{epsfig,citesort}                    \textwidth=17cm
\def\gab{g_{\alpha\beta}}
 
\def\n1{{\cal{N}}_{1i}} 
\def\PPD{\Phi^{\dagger}\Phi}
\def\CCD{X^{\dagger}X}
\def\PPC{\Phi\Phi X^{\dagger}}
\def\PPCC{(\Phi^{\dagger}\tau_i\Phi) (X^{\dagger}t_iX)} 
\def\PCCP{(\Phi^{\dagger}X)(X^{\dagger}\Phi)}
\def\CC5{\left| X^TCX\right|^2}
\def\Ppm{\phi^+\phi^-} 
\def\Pzz{\phi^0\phi^{0*}}
\def\Cpmpm{\chi^{++}\chi^{--}}
\def\Cpm{\chi^+\chi^-} 
\def\Czz{\chi^0\chi^{0*}}
\def\PpCm{\phi^+\chi^-}

\def\PzCzc{\phi^{0*}\chi^{0*}}
\def\PpCpCm{\phi^+\chi^+\chi^{--}}

\def\CpCpCm{\chi^+\chi^+\chi^{--}}

\def\mone{\mu_{1}}
\def\mtwo{\mu_{2}}
\def\lone{\lambda_{1}}
\def\ltwo{\lambda_{2}}
\def\lthree{\lambda_{3}}
\def\lfour{\lambda_{4}}
\def\lfive{\lambda_{5}}

\def\azero{a_{_0}}
\def\vone{v_{_1}}
\def\vtwo{v_{_2}}
\def\vprime{v_{_1}/\sqrt{2}}

\def\bra{\langle}
\def\ket{\rangle}

\def\cplus{c_{_+}}
\def\splus{s_{_+}}
\def\cimg{c_{_I}}
\def\simg{s_{_I}}
\def\creal{c_{_R}}
\def\sreal{s_{_R}}
\def\cw{c_{_W}}
\def\sw{s_{_W}}
\def\hpp{H^{++}}
\def\hmm{H^{--}}
\def\hhp{H^{+}}
\def\hlp{G^{+}}
\def\hhm{H^{-}}
\def\hlpm{G^{\pm}}
\def\hhpm{H^{\pm}}
\def\hlm{G^{-}}
\def\hhr{H^{0}}
\def\hlr{h^{0}}
\def\hhi{A^{0}}
\def\hli{G^{0}}
\def\futki{&:&}
\def\pmu{P_{\mu}}
\newcommand{\ba}{\begin{array}}    
\newcommand{\ea}{\end{array}}    
\newcommand{\bd}{\begin{displaymath}}    
\newcommand{\ed}{\end{displaymath}}    
\newcommand{\be}{\begin{equation}}    
\newcommand{\ee}{\end{equation}}    
\newcommand{\bea}{\begin{eqnarray}}    
\newcommand{\eea}{\end{eqnarray}}    

\def\etal{\em{et al.}}
\def\issue(#1,#2,#3){{\bf #1}, #2 (#3)} 

\def\APP(#1,#2,#3){Acta Phys.\ Polon.\ \issue(#1,#2,#3)}
\def\ARNPS(#1,#2,#3){Ann.\ Rev.\ Nucl.\ Part.\ Sci.\ \issue(#1,#2,#3)}
\def\CPC(#1,#2,#3){Comp.\ Phys.\ Comm.\ \issue(#1,#2,#3)}
\def\CIP(#1,#2,#3){Comput.\ Phys.\ \issue(#1,#2,#3)}
\def\EPJC(#1,#2,#3){Eur.\ Phys.\ J.\ C\ \issue(#1,#2,#3)}
\def\EPJD(#1,#2,#3){Eur.\ Phys.\ J. Direct\ C\ \issue(#1,#2,#3)}
\def\IEEETNS(#1,#2,#3){IEEE Trans.\ Nucl.\ Sci.\ \issue(#1,#2,#3)}
\def\IJMP(#1,#2,#3){Int.\ J.\ Mod.\ Phys. \issue(#1,#2,#3)}
\def\JHEP(#1,#2,#3){J.\ High Energy Physics \issue(#1,#2,#3)}
\def\JPG(#1,#2,#3){J.\ Phys.\ G \issue(#1,#2,#3)}
\def\MPL(#1,#2,#3){Mod.\ Phys.\ Lett.\ \issue(#1,#2,#3)}
\def\NP(#1,#2,#3){Nucl.\ Phys.\ \issue(#1,#2,#3)}
\def\NIM(#1,#2,#3){Nucl.\ Instrum.\ Meth.\ \issue(#1,#2,#3)}
\def\PL(#1,#2,#3){Phys.\ Lett.\ \issue(#1,#2,#3)}
\def\PRD(#1,#2,#3){Phys.\ Rev.\ D \issue(#1,#2,#3)}
\def\PRL(#1,#2,#3){Phys.\ Rev.\ Lett.\ \issue(#1,#2,#3)}
\def\SJNP(#1,#2,#3){Sov.\ J. Nucl.\ Phys.\ \issue(#1,#2,#3)}
\def\ZPC(#1,#2,#3){Zeit.\ Phys.\ C \issue(#1,#2,#3)}

\begin{document} 
\begin{titlepage}
\begin{flushright} 
HRI-P08-02-003\\
HRI-RECAPP-08-02\\
CU-PHYSICS/02-2008
\end{flushright} 

\vskip 5pt

\begin{center} 
{\Large {\sf\textbf {Some consequences of a Higgs
triplet}}} \\
[5mm] 
{\bf Paramita Dey ${}^a$\footnote{E-mail:
paramita@mri.ernet.in}}, 
{\bf Anirban Kundu ${}^b$\footnote{E-mail:
akphy@caluniv.ac.in}}, 
{\bf Biswarup Mukhopadhyaya
${}^a$\footnote{E-mail: biswarup@mri.ernet.in}}
\vskip 10pt 
${}^a$ {Regional Centre for Accelerator-based Particle Physics,\\
Harish-Chandra Research Institute, Chhatnag Road, Jhusi, 
Allahabad 211019, India}\\ 
${}^b$ {Department of Physics, University
of Calcutta, 92 A.P.C. Road, Kolkata 700009, India}

\normalsize
\end{center} 
\begin{abstract} 
We consider an extension of the scalar sector of the Standard Model
with a single complex Higgs triplet $X$. Such extensions are the most
economic, model-independent way of generating neutrino masses through
triplet interactions. We show that a term like $\azero\Phi\Phi X^\dag$
must be included in the most general potential of such a scenario, in
order to avoid a massless neutral physical scalar.  We also
demonstrate that $\azero$ must be real, thus ruling out any additional
source of CP-violation. We then examine the implications of this term
in the mass matrices of the singly-and doubly-charged scalar, neutral
scalar and pseudoscalar fields. We find that, for small values of
$\azero/\vtwo$, where $\vtwo$ is the triplet vev, the spectrum allows
the decay of heavier scalars into lighter ones via gauge
interactions. For large $\azero/\vtwo$, the doubly-charged,
singly-charged and neutral pseudoscalar bosons become practically
degenerate, while the even-parity neutral scalars remain considerably
lighter, thus emphasizing the possibility of decay of the
singly-charged or neutral pseudoscalar states into the neutral
scalars. Constraints from the $\rho$-parameter are used to find
nontrivial limits on the charged Higgs mass depending on $\azero$. We
also study the couplings of the various physical states in this
scenario. For small values of $|\azero|/\vtwo$, we find the lightest
neutral scalar field to be triplet-dominated, and thus having
extremely suppressed interactions with fermion as well as gauge boson
pairs.

\vskip 5pt \noindent \texttt{PACS Nos: 12.60.Fr, 14.80.Cp} \\
\texttt{Keywords: Extended scalar sector, Higgs triplet}
\end{abstract} 

\begin{flushright} 
\end{flushright} 

\pagestyle{plain}

\end{titlepage}

\setcounter{page}{1}


\setcounter{footnote}{0}
\renewcommand{\thefootnote}{\arabic{footnote}}

\section{Introduction}
\label{intro}
Even though the Standard Model (SM) of electroweak interactions has
proven to be enormously successful, it is not obvious that a single
Higgs doublet field is responsible for giving masses to the weakly
interacting vector bosons and fermions \cite{higgs}.  While fermion
masses can arise only through Yukawa couplings with Higgs doublets,
gauge bosons can acquire masses from higher representations of SU(2)
as well. Although phenomenological constraints such as that from the
$\rho$-parameter restrict the vacuum expectation values (vev) of
scalar multiplets higher than dimension 2
\cite{gvw1,gvw2,mahabharat,triplet-rho}, such multiplets are not
necessarily without phenomenological significance
\cite{akbm,triplets1,tripletferms,triplets2}. For example, Higgs
triplets can generate Majorana masses for neutrinos once $\Delta L =
2$ interactions are allowed, thereby avoiding the necessity of
right-handed neutrinos \cite{tripletnu}. Higgs triplets are also a
part of the particle spectrum of some theories attempting
stabilization of the electroweak symmetry scale, such as Little Higgs
models, even in their relatively economical forms. Higher
representations of scalars have some additional phenomenological
implications such as $WZ$ interactions of a singly charged scalar
\cite{tripletdecays}.

Extensions of the Higgs sector of the SM employing additional singlet
\cite{singlets}, doublet
\cite{deshp,doublets-th,doublets-cp,doublets-nu,doublets-collider,doublets-susy}
as well as triplet fields
\cite{gvw1,gvw2,mahabharat,triplet-rho,akbm,triplets1,tripletferms,triplets2}
have frequently been considered in the literature. Among these the
extensions involving Higgs triplets are particularly interesting,
primarily because of their capacity to generate neutrino masses
\cite{tripletnu}, as mentioned above. There are studies in this spirit
on left-right symmetric models with or without supersymmetry
\cite{tripletLR,tripsusyLR}, the Little Higgs models \cite{triplets1},
as also on situations with both complex and real scalar triplets whose
vevs are related through a custodial symmetry
\cite{gvw1,gvw2,gvw-like}.  Some of these scenarios imply rather
interesting collider signals that has been at least partially explored
in various studies
\cite{tripletdecays,tripletLEP,triplettevatron,tripletLHC,triplet-collider}.
Such studies, as also information extracted on the mass spectrum, help
not only in understanding the overall physics of electroweak symmetry
breaking (EWSB), but also in probing the scalar potential as a
specific component of the theory.

Since it is always helpful to get a model-independent perspective, we consider
here a scenario with just the added component for neutrino mass generation,
assuming {\em one doublet ($\Phi$) and one complex triplet ($X$) scalar}.  In
such a case (as opposed to one with complex as well as real triplets, with
their vev's related), the vev of the triplet must be relatively small
($\lesssim$ 12 GeV \cite{precision}) to satisfy the constraint on the
$\rho$-parameter (which translates into a constraint arising from tree-level
contributions to the electroweak precision variable $T$)
\cite{gvw1,gvw2,mahabharat,triplet-rho}. The other important oblique
parameter, namely, $S$, does not provide any serious constraint on this
scenario (including the special inputs of this study outlined below), since
the mixing of the triplet scalar with the doublet, being proportional to the
triplet vev, is small \cite{spaper}.

The salient features of this study, and the new observations arising
therefrom, are as follows:

\begin{itemize}
\item A term proportional to $\PPC$ is retained in the scalar
potential and not left out by invoking a discrete symmetry, as has
often been done in recent studies \cite{gvw1,gvw2}.

\item It is seen that, when there is no real scalar triplet, leaving
out the above trilinear term (which adds a dimension-full parameter to
the Lagrangian \cite{gavela}) implies a global $O(2)$ symmetry in the
neutral scalar sector. Giving the neutral fields vev thus results in
an additional Goldstone boson, which remains as a physical field, and
is inconsistent with experimental observations. {\em With just one
complex triplet added, such an unacceptable situation is avoided only
if the trilinear term is retained}.

\item It can be seen from very general considerations that the
coefficient of the trilinear term, must be real. Thus its introduction
does not entail any additional CP-violating phase(s), implying that
the scalar sector cannot contain any seed of CP-violation with one
doublet and one complex triplet only. As a corollary, we show that
more than one multiplet of any given kind is necessary to have
CP-violation in the scalar sector.

\item The value as well as the sign of the coefficient of the
trilinear term is subject to rather non-trivial constraints from the
electroweak symmetry breaking conditions, the requirement of the
potential bounded from below, and the absence of tachyonic modes.

\item The ordering of the scalar spectrum and the composition of the
lightest neutral scalar depend on the coefficient of the trilinear
term, as demonstrated using several benchmark values of different
parameters occurring in the model. This in turn affects the fermionic
and gauge couplings of the low-lying physical states, and restricts
the viability of different decay chains of the relatively heavier
states.
\end{itemize}

The field content and the structure of the potential have been
outlined in Section 2, where we have also discussed why such a
potential cannot lead to CP-violating effects. In Section 3, we
compute the masses of the scalar fields, and in Section 4, various
constraints on the scalar potential are discussed. Taking all these
constraints into account, we show, in Section 5, the mass spectrum as
well as the couplings of the scalar fields to fermions and gauge
bosons, where we also discuss their overall implications.  We
summarize and conclude in Section 6.

\section{The triplet Higgs model}
\label{om}
In the simple model that we consider here, the Higgs sector consists
of a complex scalar $Y = 2$ triplet $X$, along with the usual complex
$Y = 1$ doublet $\Phi$ of the SM:
\begin{eqnarray}
\Phi =\ \left(\! \begin{array}{c} \phi^{+} \\ \phi^{0}
\end{array}\!\right)\,, ~~ X =\ \left(\! \begin{array}{c} \chi^{++}
  \\ \chi^{+} \\ \chi^{0\ast} \end{array}\!\right)\,.
\label{defs}
\end{eqnarray}
We choose phase conventions for the fields such that
$\left(\chi^{++},\chi^{+},\phi^+\right)^\ast=\left(
\chi^{--},\chi^{-},\phi^-\right)$, and assign vevs to the neutral
components as follows:
\begin{eqnarray}
\bra\phi^{0}\ket=\vprime, ~~ \bra\chi^{0}\ket=\vtwo.
\label{vevs}
\end{eqnarray}
Moreover, we can use the freedom of choosing the relative phase of $X$
and $\Phi$ and align $\vprime$ and $\vtwo$ simultaneously along the
real axis without any loss of generality. The most general potential
for the scalar sector can be written as $V_{\rm total}=V_2+V_3+V_4$,
where the subscript attached to each term stands for the number of
fields occurring in it. Individually,
\begin{eqnarray}
V_2 &=& -\mone^2\left(\PPD\right)+\mtwo^2\left(\CCD\right) \\ V_3 &=&
\sqrt{3}\azero\left(\PPC\right) + {\rm h.c.} \\ V_4 &=&
\lone\left(\PPD\right)^2 + \ltwo\left(\CCD\right)^2 +
\lthree\left(\PPD\right)\left(\CCD\right) + \lfour\PPCC + \lfive\CC5,
\label{v's}
\end{eqnarray}
where
\begin{eqnarray}
C =\
\left(\! \begin{array}{ccc}
0 & 0 & 1 \\ 0 & 1 & 0 \\ 1 & 0 & 0
\end{array}\!\right)\,,
\end{eqnarray}
and $\tau_i$s and $t_i$s ($i=1$-3) are the Pauli matrices in 2 and 3
dimensions respectively. The factor of $\sqrt{3}$ in $V_3$ is taken
for later convenience.  Note that in the bilinear term $V_2$, we put a
`wrong sign' for the mass term for the doublet fields for spontaneous
symmetry breaking to take place, while we put a `correct sign' for the
mass term for the triplet fields. This is required for keeping the
triplet vev naturally small \cite{triplets1}, which in turn is needed
for avoiding large corrections to the $\rho$-parameter. This choice
ensures that the triplet vev arises only through the trilinear and
quartic terms, and can remain small without requiring the triplet
mass(es) to be below acceptable limits.

The trilinear term $V_3$ has often been neglected in the literature
\cite{gvw1,gvw2}. One way to do this is to demand the potential to be
invariant under the discrete transformations $\Phi\to - \Phi$ and
$X\to - X$. However, as we mentioned earlier, imposing such a discrete
symmetry has no definite theoretical motivation. In particular, such a
discrete symmetry in any case needs to be abandoned in the fermion
interaction terms, if one thinks in terms of neutrino Majorana masses
generated with a Higgs triplet \cite{tripletnu}. Furthermore, a term
of the form $\PPC$ bears a close analogy to one like $\ell\ell X$, as
far as gauge structure is concerned. Since the latter is an
indispensable part of the Type II seesaw mechanism for the generation
of neutrino masses, retaining the trilinear scalar term seems to be
quite natural. In addition, it also helps one in understanding the
smallness of the triplet vev, by postulating a positive mass-squared
term for the triplet, and letting it develop a vev through
doublet-triplet mixing only (second reference of \cite{triplets1}).

Keeping this term also enables us to ward off an additional
undesirable Goldstone boson, as will be shown in the next
section. Though this term could in principle have a complex
coefficient, which might indicate an extra source of CP-violation, we
shall show below that actually the parameter $\azero$ has to be real.

The third and fourth terms of $V_4$ represent two singlet states that
can be constructed out of two $\Phi$ fields and two $X$ fields. The
product ${\bf 2}\otimes{\bf 2}\otimes{\bf 3}\otimes{\bf 3}$ of SU(2)
contains two mutually orthogonal singlet combinations. Any other
combination, like the singlet coming out of $\PCCP$, can be expressed
in terms of these two singlets.

We expand the neutral components of both $X$ and $\Phi$ about their
vevs, and write
\begin{eqnarray}
\phi^0 = \frac{1}{\sqrt2}\left(\phi^{0R} + \vone + i\phi^{0I}\right), ~~ 
\chi^0 = \frac{1}{\sqrt2}\left(\chi^{0R} + \sqrt2\vtwo + i\chi^{0I}\right).
\label{zzexpns}
\end{eqnarray}
Next, using $\PPD = \left(\Ppm + \Pzz\right)$, and $\CCD =
\left(\Cpmpm + \Cpm + \Czz\right)$, we express all the terms in $V$ in
terms of the component fields:
\begin{eqnarray}
V_2 &=& -\mone^2 \left[ \Ppm +\Pzz \right] + \mtwo^2\left[
\Cpmpm + \Cpm + \Czz \right],\nonumber\\
V_3 &=& \azero \left[\phi^0\phi^0\chi^0 - \sqrt2 \phi^+\phi^0\chi^- +
  \phi^+\phi^+\chi^{--}\right] + {\rm h.c.},\nonumber\\
V_4 &=& \lone \left[ \Ppm\Ppm + \Pzz\Pzz  +
2\Ppm\Pzz \right] \nonumber\\
 &{}& + \ltwo [\Cpmpm\Cpmpm  + \Cpm\Cpm
 + \Czz\Czz + 2\Cpmpm\Czz\nonumber\\
&{}& +  2\Cpm\Czz  +
2\Cpmpm\Cpm ] \nonumber\\
&{}& + \lthree [ -\Cpmpm\Ppm  + \Cpmpm\Pzz
 + \Ppm\Cpm + \Ppm\Czz\nonumber\\
&{}& + \Cpm\Pzz  + \Pzz\Czz ] \nonumber\\
&{}& + \lfour [ \Cpmpm\Ppm + \Cpmpm\Pzz - \Ppm\Czz  + \Pzz\Czz
\nonumber\\ 
&& + \{\sqrt2\PpCm\PzCzc + \sqrt2\PpCpCm \phi^{0*} + ~{\rm h.c.}\} ]
\nonumber\\
&& + \lfive\left[ 4\Cpmpm\Czz + \Cpm\Cpm
 -2 \{\CpCpCm\chi^{0} + ~{\rm h.c.}\} \right].
\label{v4exp5}
\end{eqnarray}
In the limit $\azero = 0$, the neutral sector of the above potential
has an additional O(2) symmetry, which keeps it invariant under a
rotation in the $\left(\phi^{0R}, \phi^{0I}\right)$ and
$\left(\chi^{0R}, \chi^{0I}\right)$ planes. Thus an extra neutral
massless Goldstone boson, over and above the one arising from the
breaking of SU(2)~$\times$~U(1), arises when $\Phi$ and $X$ acquire
vevs, making the scenario phenomenologically unacceptable. This
problem can be avoided if one also introduces a real scalar triplet
\cite{gvw1,gvw2,mahabharat}. One therefore concludes that the
trilinear term $V_3$, written in terms of the dimension-full parameter
$\azero$, must be there in the potential if one has a complex triplet
and the usual doublet. The role of $\azero$ in determining the
spectrum of physical states and their coupling to fermion or gauge
boson pairs is thus of considerable importance, if one has to
understand the phenomenology of this `most economical' scenario
involving a scalar triplet.

Is there a possibility of CP violation with a possibly complex
$\azero$? The answer is in the negative, since one can always end up
with the vevs $\vone$ and $\vtwo$ aligned without any loss of
generality. All the $\lambda$s must be real, since the quartic field
combinations are self-hermitian. Even if we start with a complex
$\azero$, $V_3$ as in equation\ (\ref{v4exp5}) can be made real by
absorbing the phase in the field combination $\Phi\Phi X^\dag$, thus
removing any chance of CP violation. This is in contrast to, say, the
most general scenario with two Higgs doublets $\Phi_1$ and $\Phi_2$,
where the freedom of adjusting the relative phase between the two
doublets does not rotate away the CP-violating phases of terms of the
form $\Phi_1^\dag \Phi_2$ and $\left(\Phi_1^\dag \Phi_2\right)^2$ at
the same time. In our case, the relative phase of $\Phi$ and $X$ shows
up only in the term proportional to $\azero$, and can therefore be
adjusted to render $\azero$ real.

The above argument also applies to a model containing one doublet
$\Phi$, one complex triplet $X$ and a real triplet $\Psi$
\cite{gvw1,gvw2}.  In this case, there can be two relative phases
among the three multiplets.  At the same time, there are only two
terms in the most general potential where the relative phases may
occur explicitly, namely, those proportional to $\Phi\Phi X^{\dagger}$
and $\Phi\Phi X^{\dagger}\Psi$ \cite{deshp}. Obviously, the freedom of
the two relative phases can be used to rotate away any phases in the
coefficients of both the above terms. Thus the tree-level potential
does not allow CP-violation with Higgs doublets, complex triplets and
real triplets so long one has {\em not more than one of a particular
kind} of multiplets.

\section{Masses and couplings of the physical scalars}
\label{spectext}
We have mentioned that the trilinear term affects the mass spectrum.
This also alters the composition of the physical states, and changes
the various constraints on the potential.

Minimization of the most general potential in our scenario leads to
the following conditions:
\begin{eqnarray}
-\mone^2 + \vone^2\lone + \vtwo^2 \left[\lthree+\lfour\right] +
 2\vtwo^2\delta &=& 0\,, \nonumber\\ 
\mtwo^2 + 2\vtwo^2\ltwo + \frac{1}{2}\vone^2
 \left[\lthree+\lfour\right] + \frac12{\vone^2}\delta &=& 0\,,
\label{ewsb2}
\end{eqnarray}
where we have introduced the dimensionless parameter $\delta$, defined as
\begin{equation}
\delta = \frac{\azero}{\vtwo}\,.
\label{defdelta}
\end{equation}
To start with, the general potential considered by us has 10
parameters: two vevs, two $\mu$'s, five $\lambda$s and
$\azero$. Making use of the two potential minimization conditions the
number of independent parameters have been reduced to 8, since the two
$\mu$'s can be expressed in terms of the other parameters.

\subsection{Mass of the doubly-charged field} 
\label{plusplus}
After collecting the coefficients of $\Cpmpm$ terms from the total
potential, replacing the neutral fields with the respective vevs, and
applying the minimization conditions on the potential, we obtain
\begin{equation}
M^2_{H^{\pm\pm}} = 4\vtwo^2\lfive - \vone^2\left(\lfour +
\frac12\delta\right)\,,
\label{doublemass}
\end{equation}
To avoid tachyonic scalars, we should take either $\lambda_4$ or $\delta$ to
be negative. We will see later that the negativity of $\delta$ is forced by
the neutral pseudoscalar mass matrix.  From now, we will denote the doubly
charged mass eigenstate by $H^{\pm\pm}$, which, in this model, is identical
with $\chi^{\pm\pm}$.

\subsection{Masses of the singly-charged fields} 
\label{plusminus}
The mass-squared matrix for the singly charged fields is
\begin{eqnarray}
{\cal M}^2_{\pm}\ =-\ \left(\! \begin{array}{cc} 2(\lfour+\delta)\vtwo^2
& -(\lfour+\delta)\vone\vtwo \\ -(\lfour+\delta)\vone\vtwo 
 & \frac12(\lfour+\delta)\vone^2
\end{array}\!\right)\,,
\end{eqnarray}
whose two eigenvalues are 
\begin{eqnarray}
0,~~-\frac{1}{2}v^2\left(\lfour + \delta\right),
\label{chargeeigval}
\end{eqnarray}
where $v=\sqrt{\vone^2+4\vtwo^2}$. The respective charged scalar mass
eigenstates turn out to be
\begin{eqnarray}
\hlpm &=& \frac{\vone}{v}\phi^{\pm}+\frac{2\vtwo}{v}\chi^{\pm}\\ \hhpm
&=& \frac{-2\vtwo}{v}\phi^{\pm}+\frac{\vone}{v}\chi^{\pm}
\label{singfinal}
\end{eqnarray}
It is important to note the following points:
\begin{itemize}
\item Doublet-triplet mixing does not depend on the parameter
   $\azero$, and is also small (since $\vtwo<<\vone$).
\item The mass of $\hhpm$ can be significantly large as it depends on
   the ratio $\azero/\vtwo$, where typically $\azero$ can be $\sim 1$
   TeV and $\vtwo$ is small. Thus the effect of the trilinear term is
   mainly to push up the mass of the dominantly triplet state, without
   changing its constitution, so that this state almost decouples from
   low-energy theory for a very high value of $\azero/\vtwo$.
\item In general, the doubly charged and singly charged mass
   eigenstates are non-degenerate (even when both $\azero$ and
   $\lfive$ vanish). This is because the lifting of degeneracy through
   SU(2) breaking can be driven by $\vone$, the electroweak scale vev.
\end{itemize}

\subsection{Masses of the neutral fields}
\label{zerozero}
The neutral scalar and pseudoscalar mass matrices in this scenario are
\begin{eqnarray}
{\cal M}^{0R} =\ \left(\! \begin{array}{cc} \vone^2\lone & B \\ B
& 2\vtwo^2\ltwo - \vone^2\delta/4
\end{array}\!\right)\,,\ \ 
{\cal M}^{0I}  =\
-\delta\ \left(\! \begin{array}{cc}
2\vtwo^2 & \frac{1}{\sqrt{2}}\vone\vtwo \\
\frac{1}{\sqrt{2}}\vone\vtwo & \frac14\vone^2
\end{array}\!\right)\,.
\label{zzmatrixir}
\end{eqnarray}
where $B=\vone\vtwo\left[\lthree+\lfour+\delta\right]/\sqrt{2}$. Two
massive even-parity physical states, denoted as $\hlr$ and $\hhr$, are
obtained upon diagonalizing ${\cal M}^{0R}$. We shall comment on the
masses and compositions of these states later in this section.

${\cal M}^{0I}$, on the other hand, has one massive physical state
$\hhi$, the other state being the neutral Goldstone boson $\hli$. The
eigenvalues of ${\cal M}^{0I}$ are
\begin{eqnarray}
0, ~~-\frac{\azero}{4\vtwo}\left({\vone^2+8\vtwo^2}\right), 
\label{imeigval}
\end{eqnarray}
whence the pseudoscalar physical state emerges as
\begin{eqnarray}
\hhi &=& \frac{2\sqrt{2}\vtwo}{\sqrt{\vone^2+8\vtwo^2}}\phi^{0I} +
\frac{\vone}{\sqrt{\vone^2+8\vtwo^2}}\chi^{0I}
\label{imzzfinal}
\end{eqnarray}
with a mass-squared value given by
$-\frac12\delta\left({\vone^2+8\vtwo^2}\right)$.

The following observations can be made on the mass eigenstates in the
neutral sector:
\begin{itemize}
\item $\delta$, and hence $\azero$, must be {\em negative} to avoid a
tachyonic scalar.
\item Without the $\azero$ term, the model is plagued with the
   additional Goldstone boson $\hhi$. The root of this lies in an
   additional O(2) symmetry in the neutral sector, which is broken
   explicitly when $\azero$ in non-zero. Such a symmetry could have
   been there in the charged sector as well, but for the term
   $\PPCC$. This would have led to a massless scalar having SU(2)
   gauge couplings. Since the experimental observations on $Z$-decay
   disallows such a scalar, {\em the trilinear term driven by
   $\azero$, which breaks the O(2) explicitly, is therefore a
   necessary requirement of a model where there is just an SU(2)
   doublet and a complex triplet} (the simultaneous presence of a real
   triplet breaks this O(2), and thus a trilinear term is avoidable in
   the potential of such a model).
\item As in the charged scalar sector, doublet-triplet mixing in the
   pseudoscalar sector does not depend on the parameter $\azero$.
\end{itemize}

\section{Additional constraints on the potential}
\label{allcons}
There are some additional constraints on the remaining parameters,
arising from the demand that (a) all the physical scalars must have
non-negative mass-squared values, (b) the potential has to be bounded
from below, and (c) $V_{\rm min} < 0$ for spontaneous symmetry
breaking.

\subsection{Absence of tachyonic modes}
\label{allmass}
From the requirement that all the eigenvalues of the mass-squared
matrices should be positive, one obtains the following conditions:\\
For the doubly charged field
\begin{equation}
4\vtwo^2\lfive - \vone^2\left(\lfour +\frac12\delta\right) > 0\,,
\label{para1}
\end{equation}
for the singly charged fields
\begin{equation}
-\frac12 v^2\left(\lfour+\delta\right) > 0\,,
\label{para3}
\end{equation}
and for the physical neutral pseudoscalar field   
\begin{equation}
\azero < 0\,.
\label{para2}
\end{equation}
Since $\vtwo << v$, we can drop the first term of equation\ (\ref{para1})
as a first approximation, which in turn gives the stronger constraint
\begin{equation}
\lfour < -\frac12\delta\,.
\label{para4}
\end{equation}

\subsection{Boundedness from below}
\label{bfb}
In order to examine the boundedness of the potential from below, we
first extract the part of the potential involving neutral fields only:
\begin{eqnarray}
V &=& -\mu^2_1\left(\phi^0\phi^{0*}\right) +
\mu^2_2\left(\chi^0\chi^{0*}\right) + \lone
\left(\phi^0\phi^{0*}\right)^2 + \ltwo \left(\chi^0\chi^{0*}\right)^2
+ \lthree \left(\phi^0\phi^{0*}\right)\left(\chi^0\chi^{0*}\right) 
\nonumber \\ && + \lfour
\left(\phi^0\phi^{0*}\right)\left(\chi^0\chi^{0*}\right) + \azero
\left(\phi^0\phi^0\chi^0 + {\rm h.c.}\right)\,.
\label{eqnvmin}
\end{eqnarray}
The following conditions follow from the requirement of boundedness
from below:
\begin{enumerate}
\item Since the terms quartic in fields are dominant ones, positivity
  of $\lone$ and $\ltwo$ ensures that the potential is bounded from
  below in absence of any coupling between doublet and triplet fields.
\item The third, fourth, fifth and sixth terms of the potential
  together can be written as
\begin{eqnarray}
\left[\sqrt{\lone}\left(\phi^0\phi^{0*}\right) -
\sqrt{\ltwo}\left(\chi^0\chi^{0*}\right)\right]^2 +
\left(\lthree+\lfour+2\sqrt{\lone\ltwo}\right)
\left(\phi^0\phi^{0*}\right)\left(\chi^0\chi^{0*}\right),
\end{eqnarray}
where the first term is non-negative and vanishes in the direction
$|\phi^0|^2/|\chi^0|^2 = \sqrt{\ltwo/\lone}$.  In that case, in order
that the potential is bounded from below, one requires
\begin{eqnarray}
\left(\lthree+\lfour+2\sqrt{\lone\ltwo}\right)>0\,.
\end{eqnarray}
\item One can still have potentially dangerous directions in which the
  combined contribution of the quartic terms vanishes. However, a
  delineation of conditions ensuing from this requires the computation
  of higher-order corrections to the potential, which can yield
  conditions on $\azero$ for either a potential bounded from below, or
  a false vacuum whose lifetime exceeds the age of the universe
  \cite{falsevacuum}.
\end{enumerate}

\subsection{$V_{\rm min}<0$}
\label{vminltz}
To implement this condition, we start from the total potential keeping
neutral fields only and replace them by their vevs, to obtain
\begin{eqnarray}
f\left(\vone,\vtwo\right)= -\frac{1}{2}\mone^2 \vone^2 +
\mtwo^2\vtwo^2 + \azero\vone^2\vtwo + \frac{1}{4}\lone\vone^4 +
\ltwo\vtwo^4 +\frac{1}{2}\left[\lthree+\lfour\right]\vone^2\vtwo^2.
\end{eqnarray}
Substituting $\mone^2$ and $\mtwo^2$ from equation\ (\ref{ewsb2}) in
$f\left(\vone,\vtwo\right)$, we get
\begin{eqnarray}
f_{\rm min}= -\frac{1}{4}\lone\vone^4-\ltwo\vtwo^4 -
\frac{1}{2}(\lthree+\lfour+\delta)\vone^2\vtwo^2\,.
\end{eqnarray}
The negativity of the potential at the minimum thus requires
\begin{eqnarray}
\left|\delta\right| <
\left(\frac12\lone\frac{\vone^2}{\vtwo^2} +
2\ltwo\frac{\vtwo^2}{\vone^2} + \lthree+\lfour \right),
\end{eqnarray}
which, to the leading order, can be written as,
\begin{eqnarray}
\left|\delta\right| < \frac12\lone \left(\frac{\vone}{\vtwo}\right)^2\,.
\label{a0upper}
\end{eqnarray}
Two more conditions for $V_{\rm min}$ come from the requirements
$\partial^2 f/\partial\vone^2>0$ and $\partial^2f/\partial\vtwo^2>0$,
which can simply be written as ${\cal M}^{0R}_{11}, {\cal M}^{0R}_{22}
> 0$.

We observe from equation\ (\ref{a0upper}) that not only does $\azero$ have
to be negative, but is also bounded above, given the constraint on the
magnitude of $\vtwo$.  However, the upper limit on $\azero$ can be way
above the TeV scale if $\vtwo$ is very small.  We shall see later that
the magnitude of $\azero$ is subject to further constraints, which
avoid this possibility.

The allowed region in the parameter space of this model has to satisfy
all of the above conditions. They have been taken into account in the
numerical studies on the mass spectrum and the strengths of various
couplings, reported in the next section.

\section{Some numerical results}
\label{numerics}
In the following two sections we talk about the numerical values of
masses of the physical scalar fields as well as their couplings to the
fermions and gauge bosons, and discuss their overall implications.

\subsection{Mass spectrum: numerical values}
\label{masspec}
 \begin{figure}[htbp]
    \begin{center}
      {\resizebox{8cm}{!}{\includegraphics{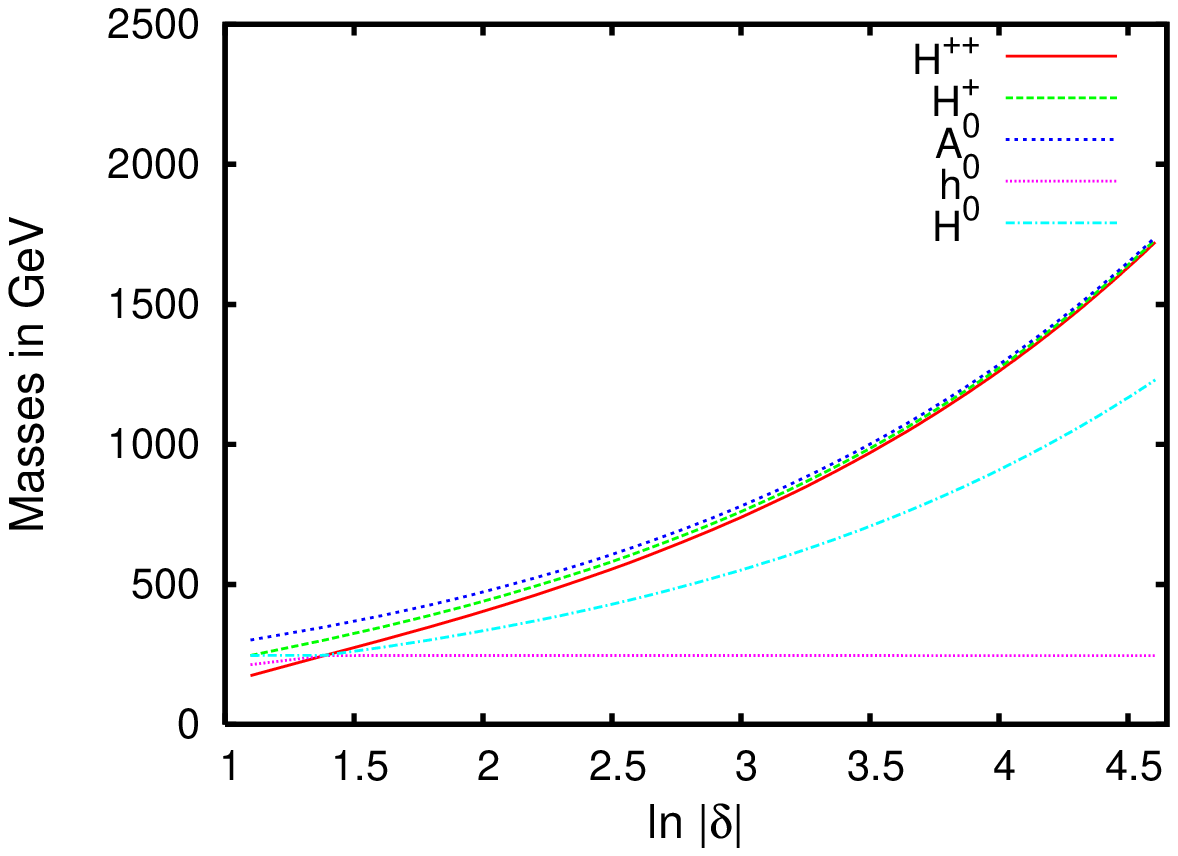}}}
      \hspace*{-2.2cm}
      {\resizebox{8cm}{!}{\includegraphics{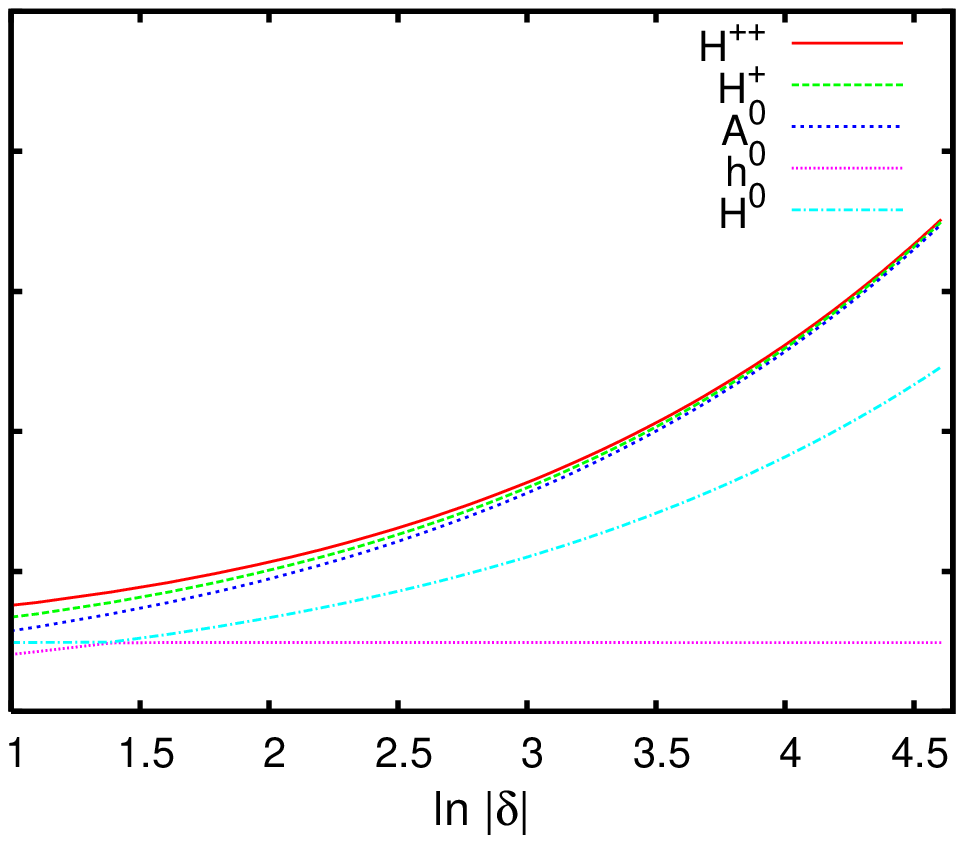}}}
      \vspace*{-0.8cm}\begin{center} {\bf (a)}\end{center}
      {\resizebox{8cm}{!}{\includegraphics{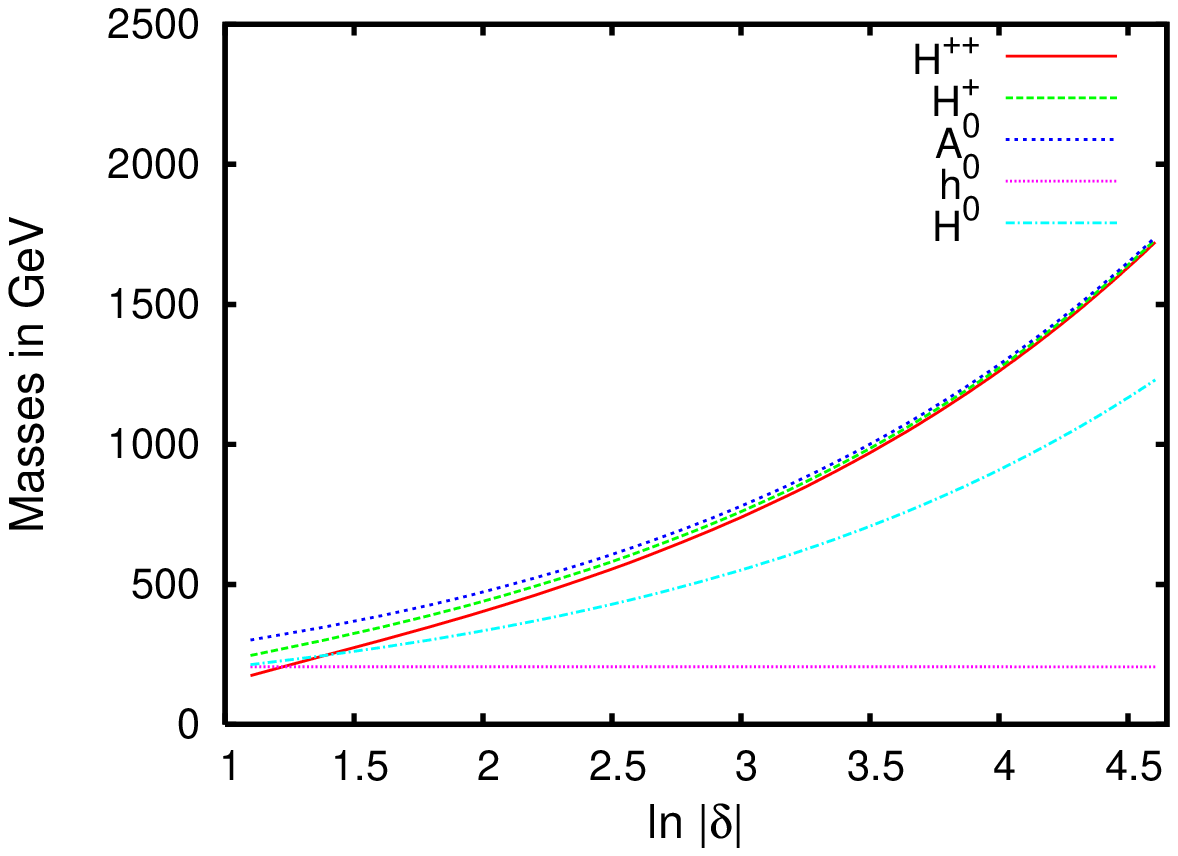}}}
      \hspace*{-2.2cm}
      {\resizebox{8cm}{!}{\includegraphics{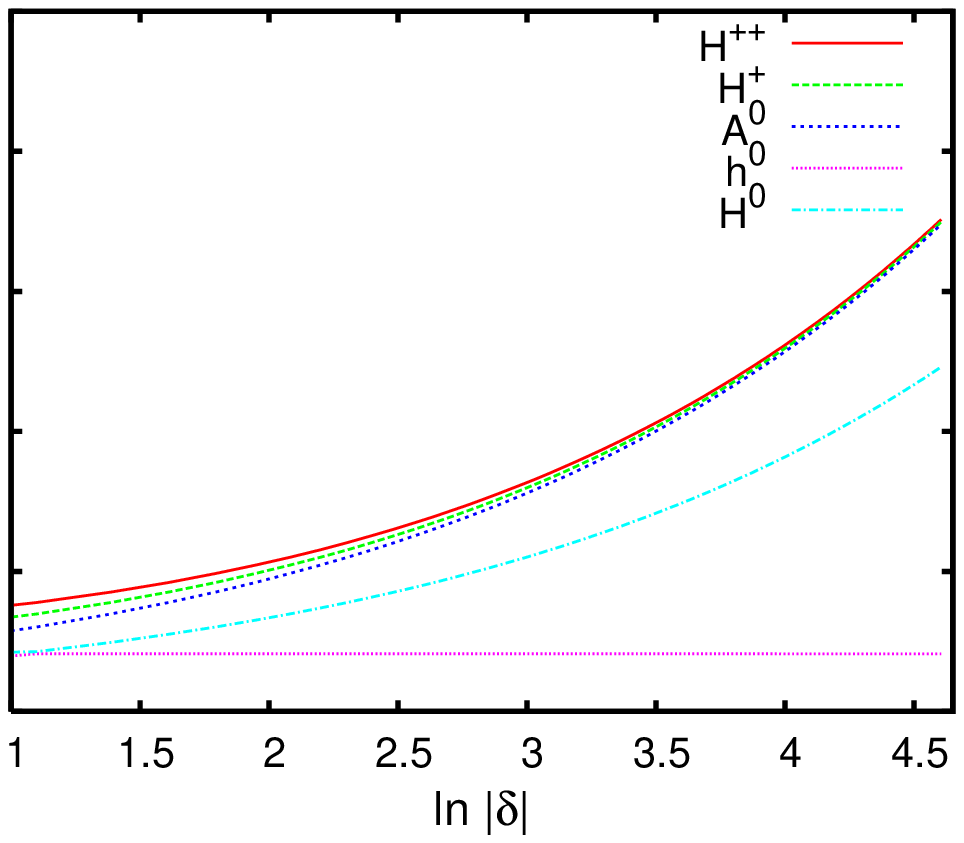}}}
      \vspace*{-0.8cm}\begin{center} {\bf (b)}\end{center}
      \label{figmass}
      \caption{Masses of the scalar physical states as functions of
        ${\rm ln}|\delta|$ ($|\delta|=|\azero/\vtwo|$). Top left
        panel: $\lambda_{i}~(i=1-5)~=1$, $\vtwo=1$ GeV, and $|\delta|$
        varies between 1 and 100, resulting in
        $M_{\hhi}>M_{\hhp}>M_{\hpp}$. Top right panel: identical with
        left but only $\lfour=-1$, resulting in
        $M_{\hhi}<M_{\hhp}<M_{\hpp}$. Bottom left panel: $\lone=0.7$,
        $\lambda_{i}~(i=2-5)~=1$, $\vtwo=1$ GeV, and $|\delta|$ varies
        between 1 and 100, resulting in
        $M_{\hhi}>M_{\hhp}>M_{\hpp}$. Bottom right panel: identical
        with bottom left but only $\lfour=-1$, resulting in
        $M_{\hhi}<M_{\hhp}<M_{\hpp}$. In all these figures the lighter
        neutral scalar $\hlr$ remains triplet-dominated below the
        cross-over point, around ${\rm ln}|\delta| \simeq 1.39$ for
        top left and top right panels and around ${\rm ln}|\delta|
        \simeq 1.03$ for bottom left and bottom right panels, beyond
        which it is doublet-dominated. It is just the opposite for
        $\hhr$.}
\end{center}
\end{figure}
Let us first remind ourselves of the roles played by various parameters in the
scalar potential in determining the mass spectrum of the model. A clear idea
of this is obtained from equations \ (\ref{doublemass}), (\ref{chargeeigval}),
(\ref{zzmatrixir}), and (\ref{imeigval}). On eliminating the mass parameters
$\mone$ and $\mtwo$ from the EWSB conditions equation\ (\ref{ewsb2}), the
physical masses are completely determined by the two vevs $\vone$ and $\vtwo$,
$\lambda_i$ ($i$ = 1-5), as well as by the dimensionless quantity
$\delta$. The scale of the doublet-dominated neutral scalar is set by
$\lone$. $\ltwo$ affects the masses at the level of $\vtwo^2$ only, while
$\lthree$ does not appear in the expressions for masses, after using the EWSB
conditions. $\lfive$ only affects the doubly charged scalar masses to order
$\vtwo^2$.

For small $|\delta|$, the masses of the states $\hhp$ and $\hpp$
depend on $\lfour$ and $\delta$, while the mass of $\hhi$ depends on
$\delta$ alone (apart from the vevs). In addition, the requirement of
making the quartic terms gauge invariant makes the different mass
matrices dependent on the SU(2) Clebsch-Gordan coefficients. The mass
expressions clearly show that for positive $\lfour$,
$M_{\hhi}>M_{\hhp}>M_{\hpp}$, while for negative $\lfour$,
$M_{\hhi}<M_{\hhp}<M_{\hpp}$.  This can also be seen in figure 1(a),
where we have varied the ratio $|\delta|$ between 1 and 100, $\vtwo$
has been set at $1~{\rm GeV}$, and all the other $\lambda$s have been
fixed at $1$. For small $|\delta|$, there is substantial separation in
the masses of $\hhi$, $\hhp$ and $\hpp$. This has interesting
implications from the viewpoint of accelerator phenomenology, since
the heavier scalars can decay into the lighter ones via gauge
couplings in this situation, as for example $\hpp\to\hhp W^+$,
$\hhp\to\hhr W^+$, $\hhp\to\hlr W^+$ and the like.

Figure 1(a) also shows that $\hlr$ can become very light for small
$\delta$, for some specific combination of parameters such as
$\lthree=-\lfour$. In such a case, a very light, triplet dominated
neutral scalar may exist, evading the limit from $e^+e^-\to
Z\hlr$. Such a scalar, however, can still be probed, for example,
through $\hlr\hhi$ production in $e^+e^-$ collision, or via the
$W^+W^-\hlr\hlr$ coupling at the LHC (in the gauge boson fusion
channel). We do not discuss the phenomenology of such a situation
here, especially because the stability of such a small $\hlr$ mass
against radiative corrections is yet to be demonstrated.  It is also
to be noted that in the limit $\azero \rightarrow 0$, $M_{\hhi}$
approaches zero, because of the emergence of the global O(2) symmetry
in the neutral sector, as discussed in the subsection \ref{zerozero}.

As $|\delta|$ grows, the $\hhi$, $\hhp$, $\hpp$ states become
degenerate, because the corresponding mass eigenvalues are controlled
more and more by the term driven by $\azero$. However, the
increasingly common value of these three masses stays above those of
the heavier CP-even neutral scalars $\hhr$ and $\hlr$. Again, this
creates a lot of opportunity for the heavy scalars being produced from
the charged ones through gauge interactions. All the above features
remain identical for figure 1(b), for which the parameters are set
identical to those in figure 1(a) except for $\lone=0.7$ (the
implication of which is explained later in this section).

One should remember that the requirement of neutrino mass generation
often leads to the choice of a much smaller $\vtwo$ than 1 GeV
\cite{tripletnu}. While the mass eigenvalues depend only upon
$\delta=|\azero|/\vtwo$, it should be noted that a value of $\vtwo$ as
small as $1~{\rm eV}$ will allow very large values of $\delta$, if
$\azero$ is around the electroweak scale. An extrapolation of the
plots in figure 1 reveals that, in such a case, one has an
increasingly wide separation of the neutral scalars with the
pseudoscalar as well as the singly-and doubly-charged states. The
possibility of the heavier states decaying dominantly into the lighter
ones gets further accentuated in such a situation. However, since the
physical states are of increasingly `pure' doublets and triplets in
this region, such decay is mostly confined to the channels $\hhp \to
W^+ \hhr$ and $\hhi \to Z \hlr$.
 \begin{figure}[htbp]
    \begin{center}
      {\resizebox{8cm}{!}{\includegraphics{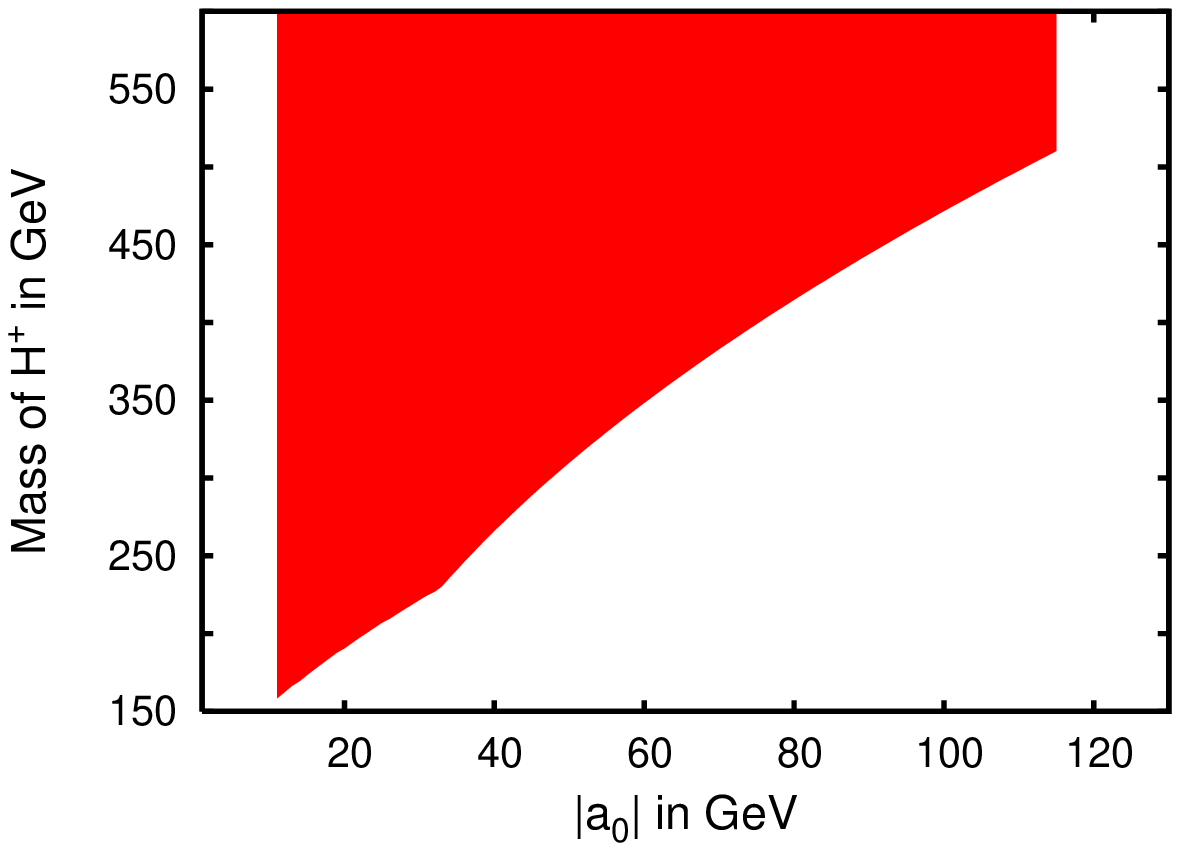}}}
      \hspace*{-1.0cm}
      {\resizebox{8cm}{!}{\includegraphics{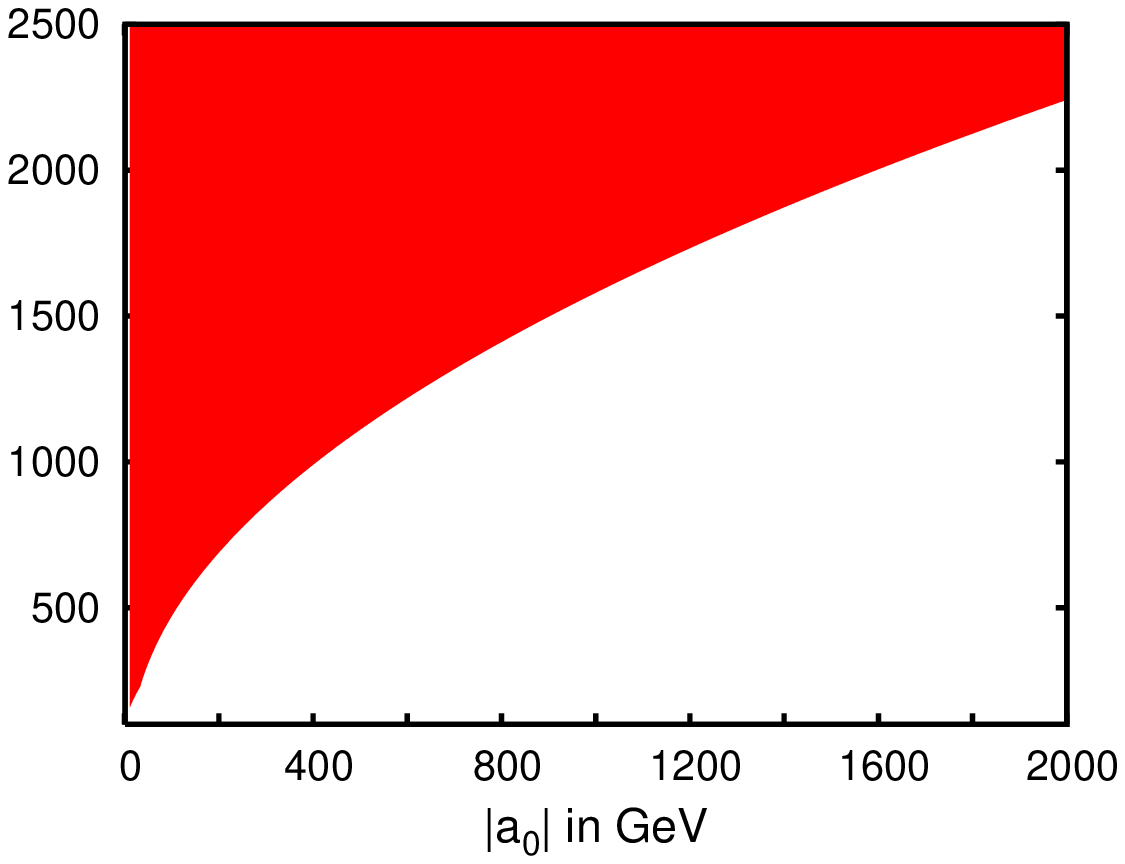}}}
      \caption{The allowed regions of $M_{\hhp}-\azero$ space, for
      $\vtwo<12~{\rm GeV}$ (coming from tree-level upper limit on the
      $\rho$-parameter), and $M_{\hlr}=120$ GeV (left panel) and
      $M_{\hlr}=400$ GeV (right panel), are shown by the shaded region marked
      by red-color. The constraints restricting this parameter space also
      include those coming from LEP results as well as from the scalar
      potential (see section \ref{allcons})}. 
\label{contest}
\end{center}
\end{figure}

Although the physical charged Higgs mass $M_{\hhp}$ for a given $M_{\hlr}$
depends on $\lfour$ and $\delta$, it is still possible to constrain regions in
the $M_{\hhp}-\azero$ space from precision data. Such constraint comes
essentially from the tree-level upper limit on the $\rho$-parameter, which in
turn implies an upper limit on $\vtwo$ \cite{precision}. Using this upper
limit (as obtained from the oblique parameter $T$), and at the same time
varying $\lfour$ over a range of admissible values (see equation
(\ref{chargeeigval})), one thus obtains a minimum value of $M_{\hhp}$ for
every $\azero$. The parameter space is further restricted by LEP results, and
the constraints on the potential discussed in section \ref{allcons} (see in
particular, equation (\ref{a0upper})). Note that such constraints may differ
for different values of $M_{\hlr}$ (decided by $\lone$), even if one scans
over the entire allowed range of other $\lambda$'s. The consequently allowed
regions of the $M_{\hhp}-\azero$ space, for $\vtwo<12~{\rm GeV}$ and two
values of $M_{\hlr}$ (120 GeV and 400 GeV) are shown in figure 2 (left and
right panels respectively). It is evident from these figures that allowed
lower limits on $M_{\hhp}$ can be significantly different from what they would
have been for $\azero=0$\footnote{As discussed in section \ref{intro}, one
gets relatively weaker constraints from loop-induced effects. Also, the
constraints on the potential may change on considering loop effects, which are
tentatively assumed to be small, since the triplet states have no coupling
with heavy quarks.}. It should be noted that a minimum value of $\azero$
is obtained; this is because the mass of the `physical' pseudoscalar state
is proportional to  $\sqrt{|\azero|}$, a fact connected with the appearance
of a glodstine boson in the limit of vanishing  $\azero$. Here we have assumed a
lower limit of 100 GeV on the pseudosalar mass, which leads to a minimum
allowed value of  $\azero$. The minimum value pertains to both panels
in the figure, though the larger range of $\azero$  in the right-hand
panel makes it nearly invisible.

Also note in both the left and the right panels of figure 1 that the
lighter neutral scalar $\hlr$ remains triplet-dominated below a
cross-over point decided by $|\delta|$, beyond which it is
doublet-dominated. For the parameter values we have chosen, this is
around $|\delta|\sim 4$ for figure 1(a), and 2.8 for figure
1(b). These cross-overs are more clearly presented in figure 3, where
we have shown the composition of $\hlr$ and $\hhr$ in terms of the
probability of it being an SU(2) doublet, as a function of $|\delta|$
for $\lfour=1$. This can be easily understood from a first-order
approximation of ${\cal M}^{0R}$ in equation (\ref{zzmatrixir}), where
we drop the relatively smaller off-diagonal terms and the $\vtwo^2$
dependent term of ${\cal M}^{0R}_{22}$. Since, for figures 3(a) for
example, we have taken $\lone=1$, one has ${\cal M}^{0R}_{11} > {\cal
M}^{0R}_{22}$ as long as $|\delta| < 4$. Thus the lighter neutral
scalar will be triplet-dominated, its mass-squared value being
approximately given by ${\cal M}^{0R}_{22}$. For both the plots of
figure 3(a), beyond this cross-over, ${\cal M}^{0R}_{22} > {\cal
M}^{0R}_{11}$, and the lighter neutral scalar becomes
doublet-dominated with ${\cal M}^{0R}_{11}$ as its approximate
mass-squared value, and its mass no longer depends on the precise
value of $\azero$. Following similar reasons, for both the plots of
figure 3(b), where we have taken $\lone=0.7$, the cross-over takes
place around $|\delta| \sim 2.8$.

Note that for small values of $\vtwo$ (see the plots in the right
panel of figure 3 where $\vtwo=1~{\rm eV}$), this cross-over is less
smooth compared to the large $\vtwo$ cases (plots in the left panel of
figure 3). This is expected since, for small $\vtwo$, the first-hand
approximation that we have made above is more exact, making ${\cal
M}^{0R}$ almost diagonal. Thus, the transition from ${\cal
M}^{0R}_{11} > {\cal M}^{0R}_{22}$ to ${\cal M}^{0R}_{22} > {\cal
M}^{0R}_{11}$ is rather sharp at $|\delta|=4\lone$ (see equation
(\ref{zzmatrixir})). The very small value of $\vtwo$ makes the
off-diagonal terms of ${\cal M}^{0R}$ inconsequential in rendering the
transition somewhat gradual.

 \begin{figure}[htbp]
    \begin{center}
      {\resizebox{8cm}{!}{\includegraphics{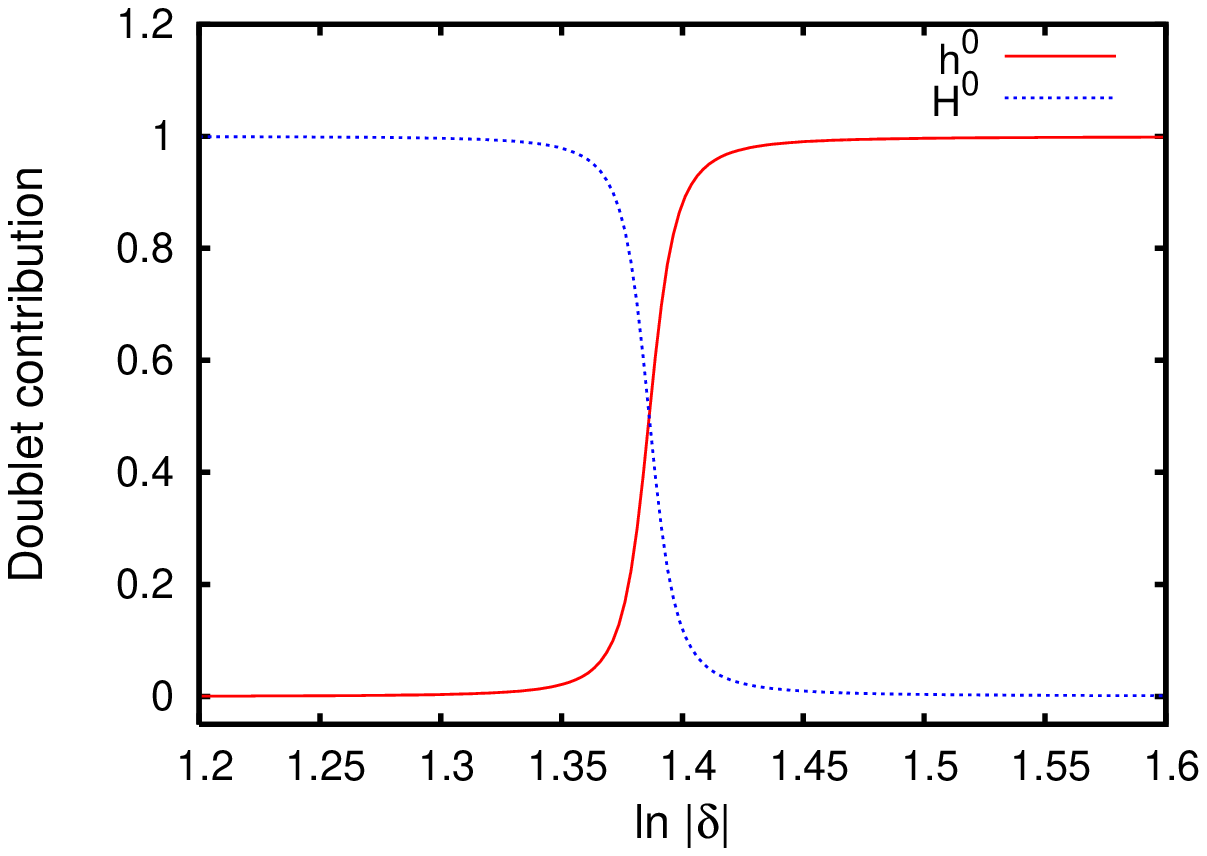}}}
      \hspace*{-1.8cm}
      {\resizebox{8cm}{!}{\includegraphics{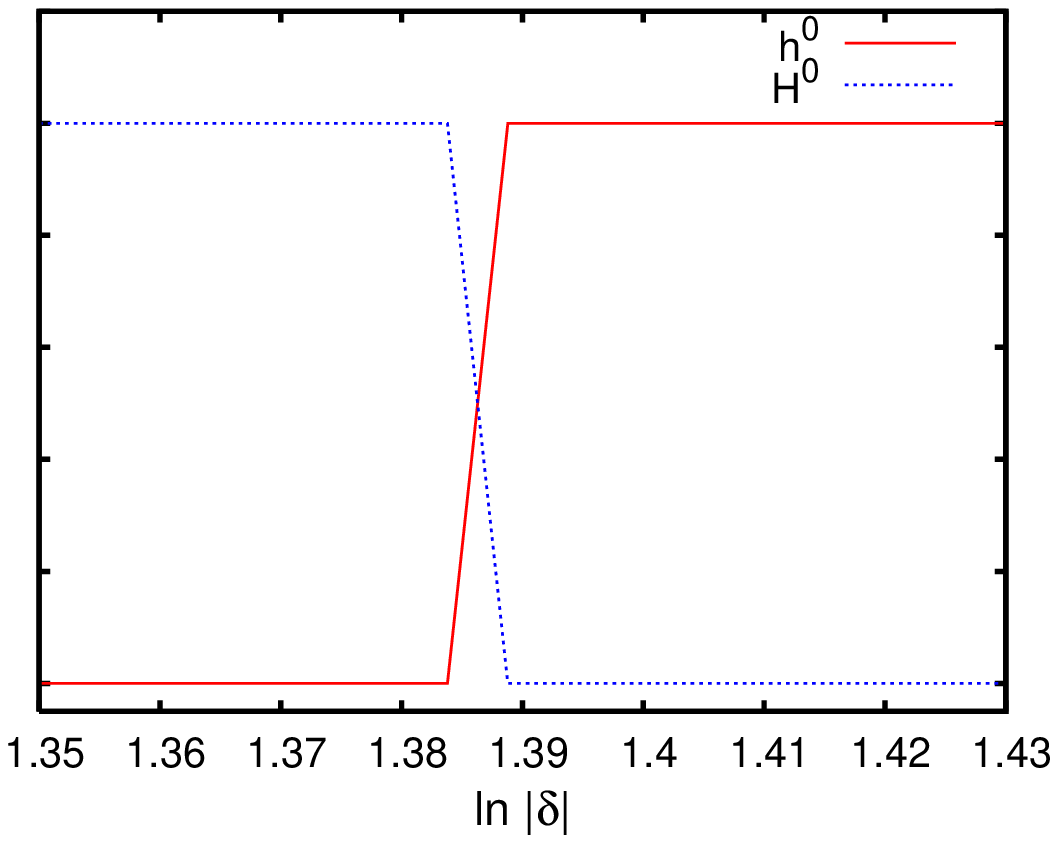}}}
      \vspace*{-0.8cm}\begin{center} {\bf (a)}\end{center}
      {\resizebox{8cm}{!}{\includegraphics{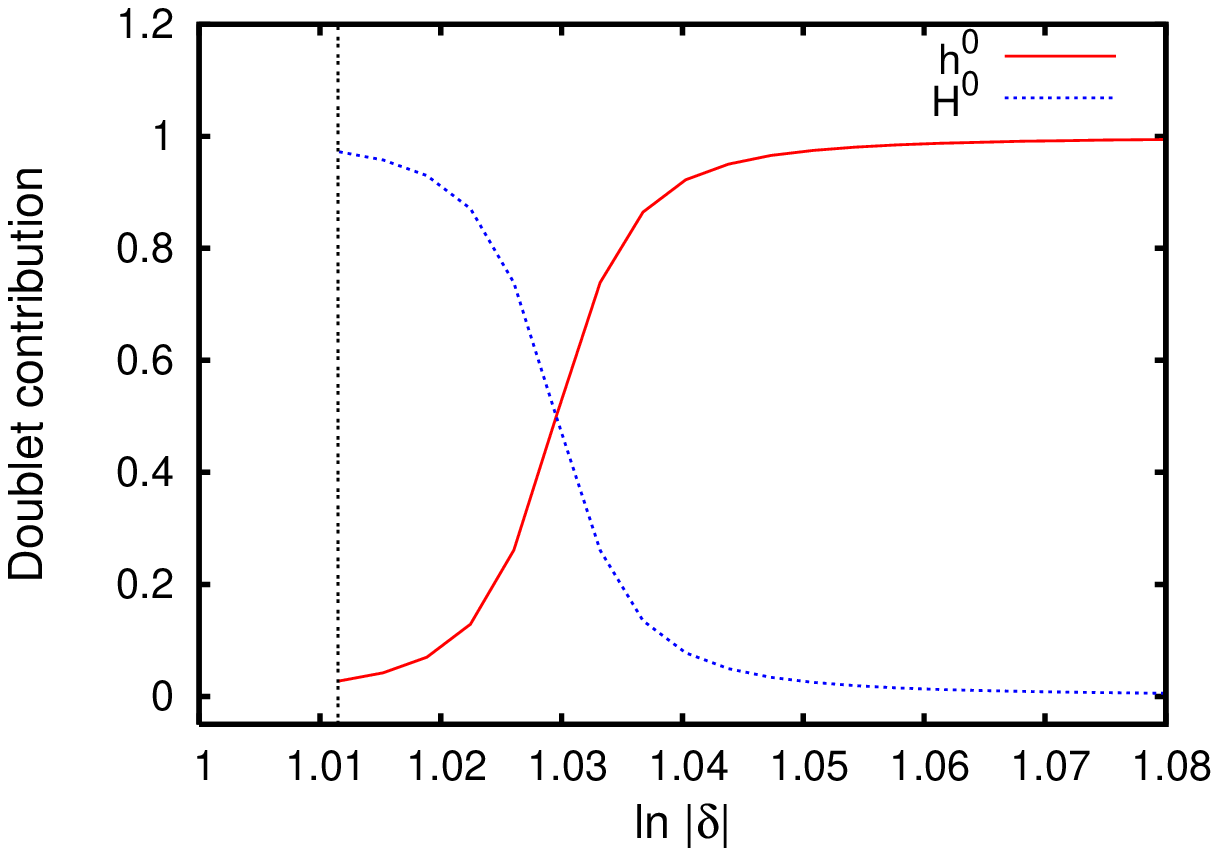}}}
      \hspace*{-1.8cm}
      {\resizebox{8cm}{!}{\includegraphics{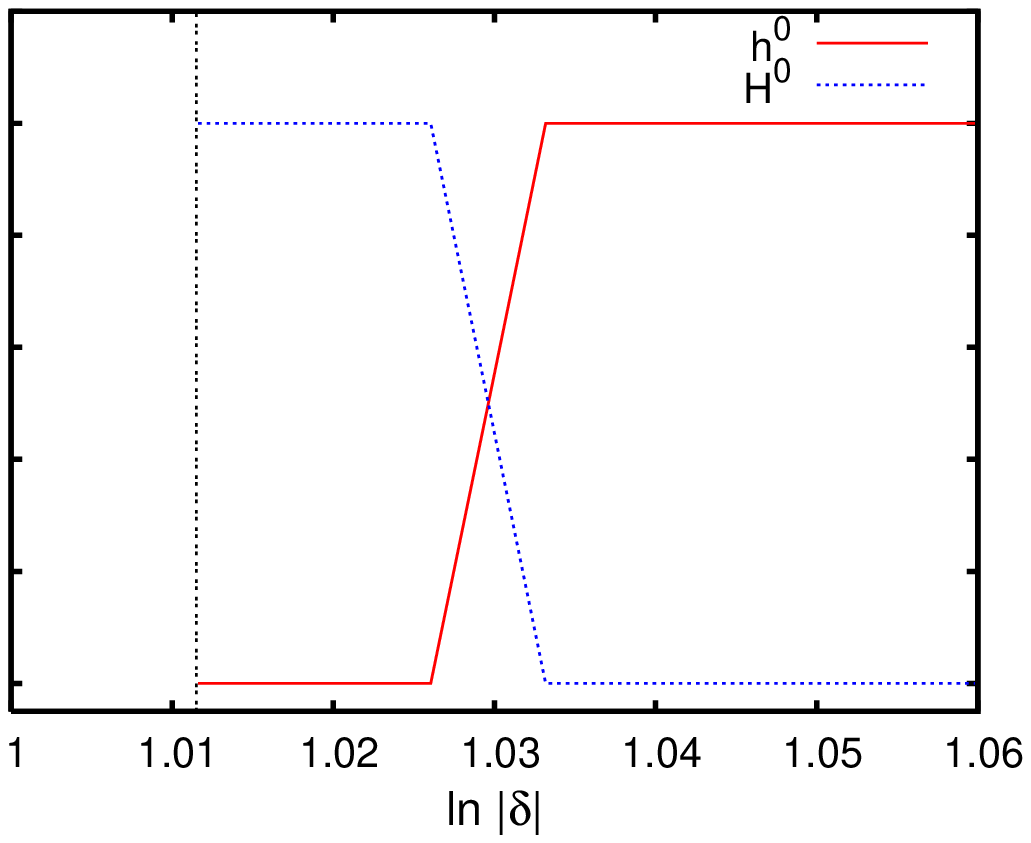}}}
      \vspace*{-0.8cm}\begin{center} {\bf (b)}\end{center}
      \caption{The doublet compositions of $\hlr$ and $\hhr$ as
    functions of ${\rm ln} |\delta|$. Top left panel:
    $\lambda_{i}~(i=1-5)~=1$ and $\vtwo=1$ GeV. Top right panel:
    identical with top left but only $\vtwo=1$ eV. Bottom left panel:
    identical with top left but only $\lone=0.7$. Bottom right panel:
    identical with top left but $\vtwo=1$ eV, $\lone=0.7$. For all
    cases, $\hlr$ remains triplet-dominated below the cross-over
    point, ${\rm ln}|\delta| \simeq 1.39$ for top panels and ${\rm
    ln}|\delta| \simeq 1.03$ for bottom panels, beyond which it is
    doublet-dominated. Just the opposite happens to $\hhr$. The
    regions on the left of the vertical line shown in the figures of
    2(b) are disallowed, by the chosen lower limits on the scalar
    masses, such as, $M_{\hpp}>150~{\rm GeV}$,
    $M_{\hhp},~M_{\hhi},~M_{\hhr}>100~{\rm GeV}$ and
    $M_{\hlr}>115~{\rm GeV}$.}
\label{compos}
\end{center}
\end{figure}

\subsection{Fermion and gauge-boson pair couplings} 
\label{cplngs}
Among the various possible fermion and gauge-boson pair couplings with the
physical scalar states, only the couplings with neutral scalars $\hhr$ and
$\hlr$ show dependence on $\azero$ (see the Appendix for a complete list of
all fermion/gauge-scalar couplings of this model). As is obvious from the
plots in figure 4, all of $t$, $b$, $W^{\pm}$ and $Z$ pair couplings show the
effect of the cross-over in the composition of $\hhr$ and $\hlr$ discussed in
the previous section. These plots have been generated for $|\delta|$ in the
range 1-6, $\vtwo$ set at 1 GeV, and the $\lambda_{i}$ ($i=2-5$) fixed at $1$,
$\lone$ has been set at $1$ for the left panel, and at $0.7$ for the right
panel. We do not show the corresponding plots for other benchmark points here
(i.\ e.\ small $\vtwo$, negative $\lfour$), since they {\it do not} add
additional information to the general observations made below.

For higher values of $|\delta|$, including the regions of very small
$\vtwo$ suggested by neutrino mass values, the lighter state $\hlr$ remains
overwhelmingly doublet-dominated, and its signatures are identical to that of
the SM Higgs boson.
 \begin{figure}[htbp]
    \begin{center}
      {\resizebox{8cm}{!}{\includegraphics{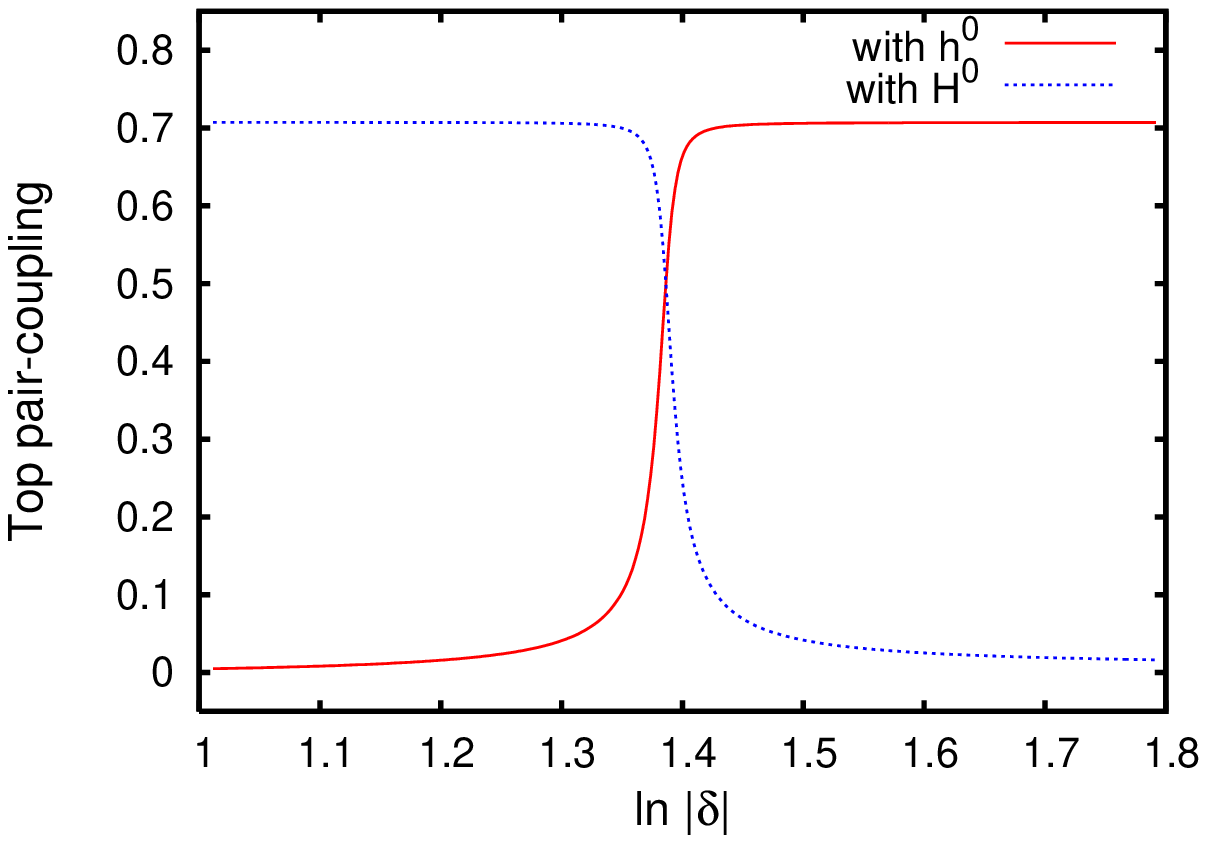}}}
      \hspace*{-1.8cm}
      {\resizebox{8cm}{!}{\includegraphics{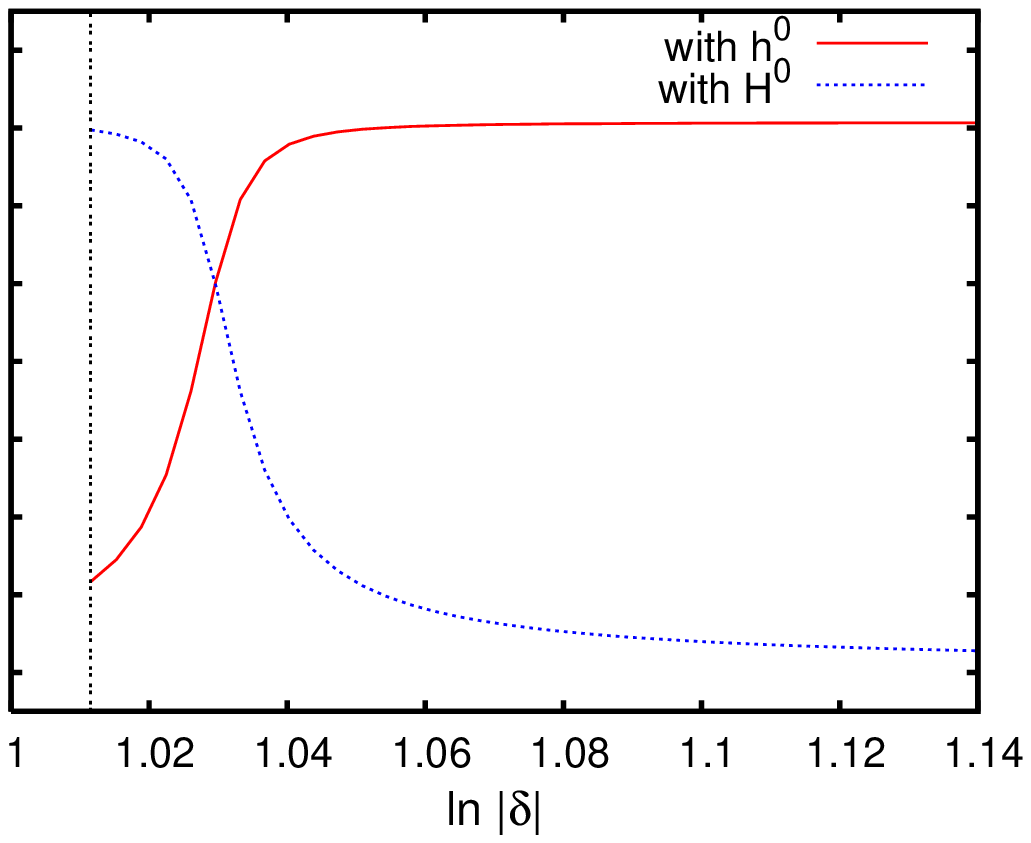}}}
      \vspace*{-0.8cm}\begin{center} {\bf (a)}\end{center}
      {\resizebox{8cm}{!}{\includegraphics{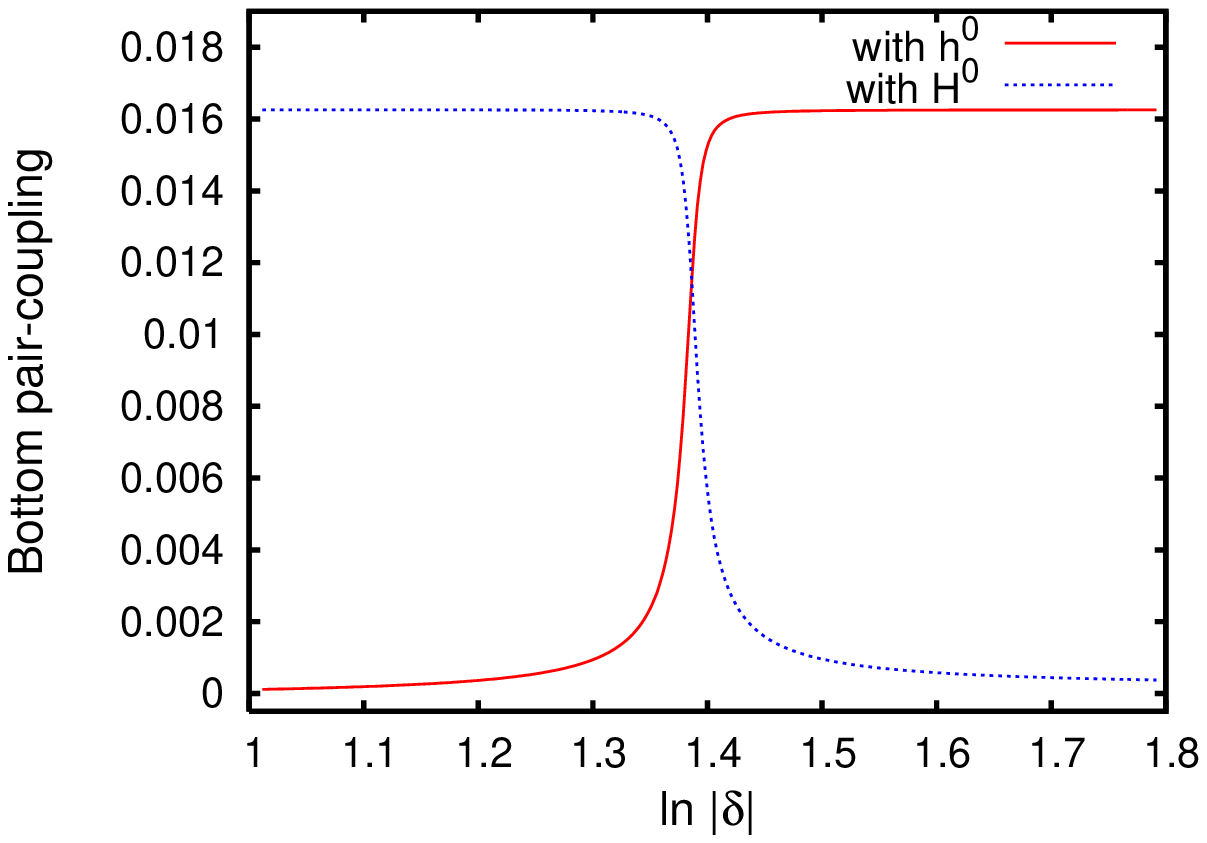}}}
      \hspace*{-2.1cm}
      {\resizebox{8cm}{!}{\includegraphics{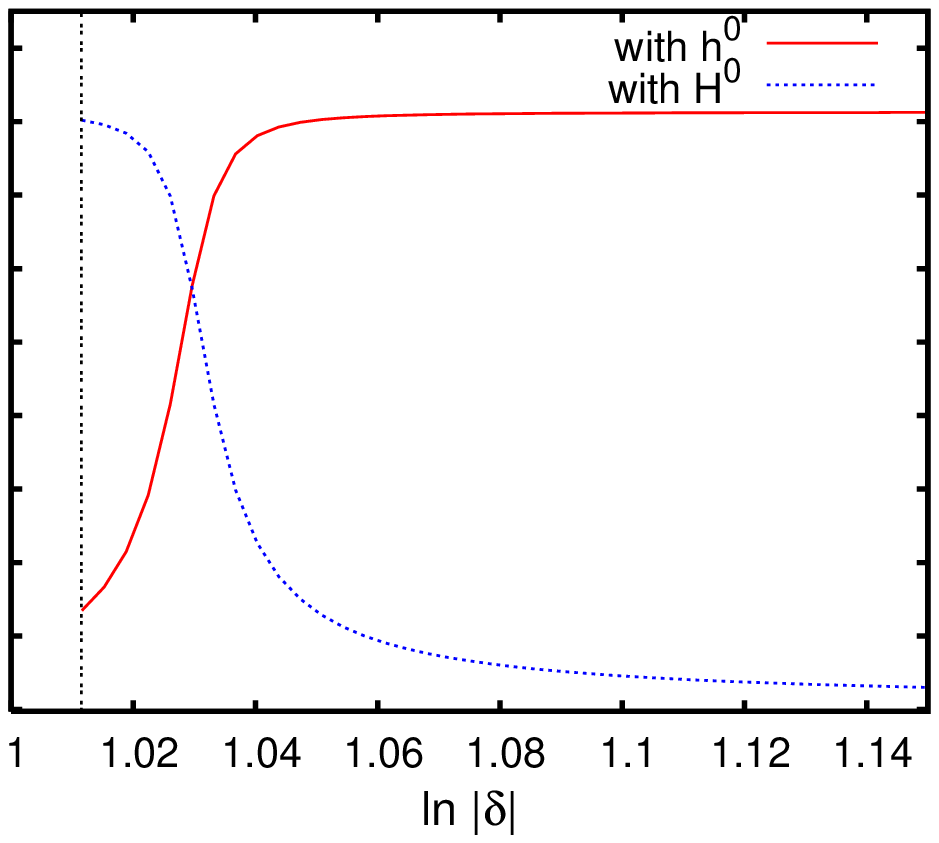}}}
      \vspace*{-0.8cm}\begin{center} {\bf (b)}\end{center}
      {\resizebox{8cm}{!}{\includegraphics{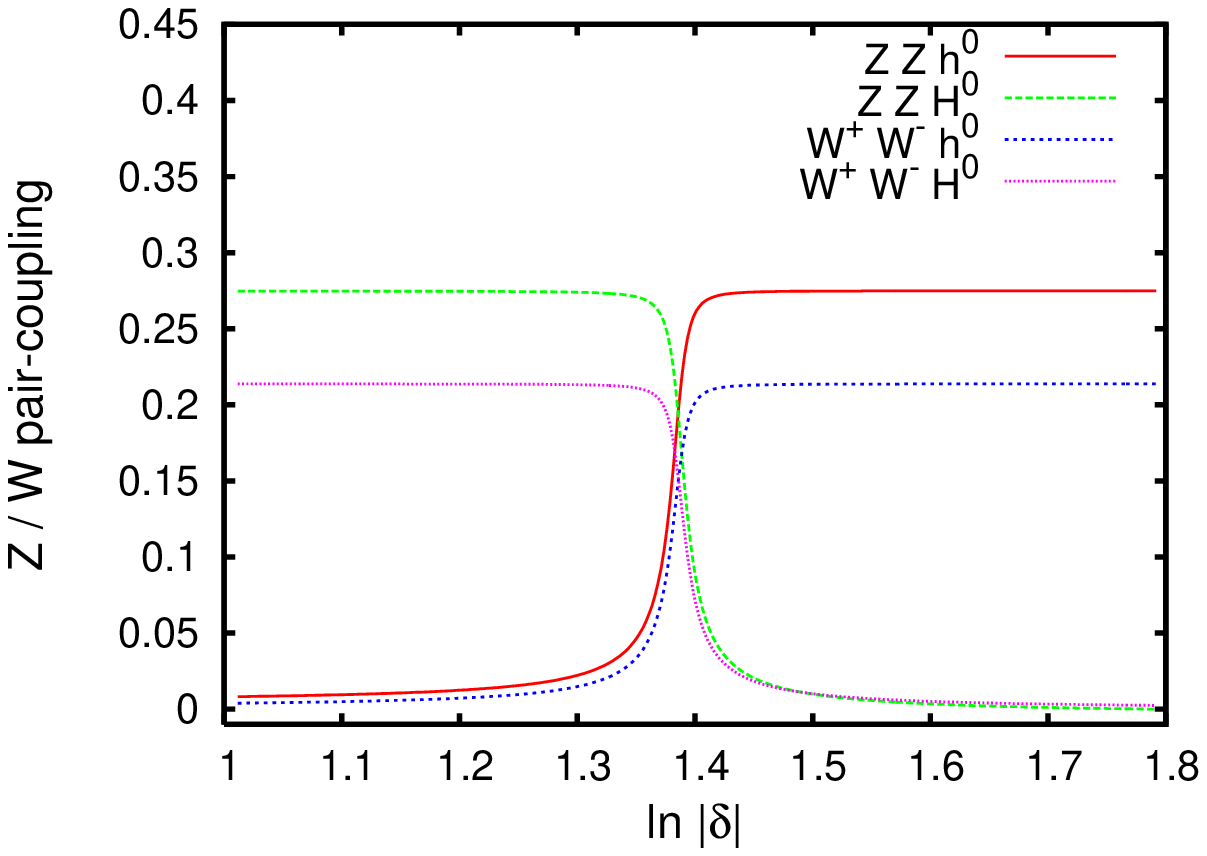}}}
      \hspace*{-2.0cm}
      {\resizebox{8cm}{!}{\includegraphics{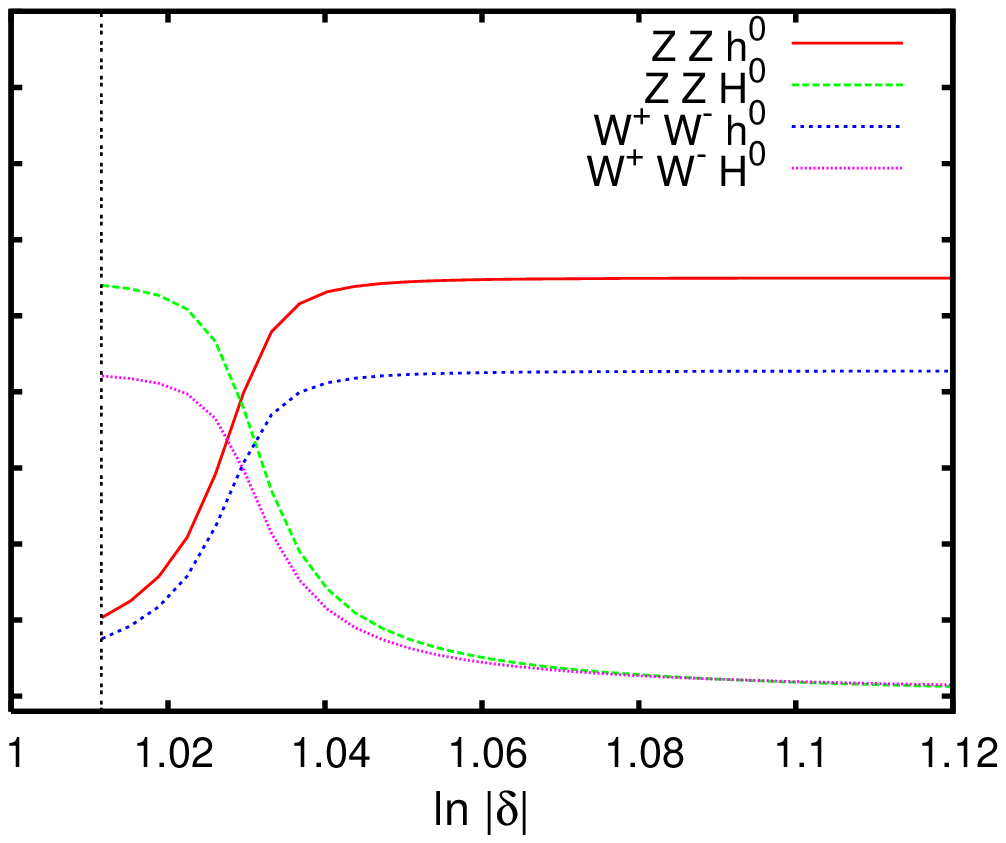}}}
      \vspace*{-0.8cm}\begin{center} {\bf (c)}\end{center}
      \label{wzfig}
      \caption{$t$, $b$, $W^{\pm}$ and $Z$ pair couplings with $\hhr$
        and $\hlr$ against ${\rm ln}|\delta|$, with
        $\lambda_i~(i=1-5)$ =1 and $\vtwo=1$ GeV (in the left panel)
        and $\lone=0.7$, $\lambda_i~(i=2-5)$ =1 and $\vtwo=1$ GeV (in
        the right panel). Below the cross-over point, ${\rm
        ln}|\delta| \simeq 1.39$ for left panel and ${\rm ln}|\delta|
        \simeq 1.03$ for right panel, interactions of $\hlr$ are
        suppressed, relaxing its experimental mass limit. The regions
        on the left of the vertical line shown in the figures in the
        right panel are disallowed, by the chosen lower limits on the
        scalar masses, such as, $M_{\hpp}>150~{\rm GeV}$,
        $M_{\hhp},~M_{\hhi},~M_{\hhr}>100~{\rm GeV}$ and
        $M_{\hlr}>115~{\rm GeV}$.}
\end{center}
\end{figure}
A somewhat more striking feature, however, presents itself below the
cross-over region ({\em i.\ e.\ }, for $|\delta|\le 4$ for the left
panel and $|\delta|\le 2.8$ for the right panel of figure 4). Although
such an $\azero$ is on the lower side of the electroweak scale, it is
still allowed, perhaps with a mild degree of fine-tuning. The
interesting point to note is that $\hlr$ is dominantly a triplet here,
and, as seen from the left panels of figures 4(a), 4(b) and 4(c), it
has rather suppressed interactions with both fermion and gauge boson
pairs. As a result, its production in all of the usually expected
channels will be suppressed. At the LHC, the gluon-gluon fusion
channel will suffer due to the feeble coupling of the lightest scalar
to the top quark in the loop, while the feeble character of gauge
boson pair coupling will suppress production via gauge boson fusion.
(However, $\hhr$ may be produced, depending on its mass.)  In $e^+
e^-$ collision, too, both the processes of associated production of
$\hlr$ with $Z$ as well as gauge boson fusion will undergo suppression
in an identical manner. Consequently, the experimental limit on the
mass of such a scalar is much more relaxed than in the case of a Higgs
doublet \cite{tripletLEP,triplettevatron,tripletLHC,triplet-collider}.

\section{Summary and conclusions}
\label{summary}
We have considered the inclusion of a single complex Higgs triplet $X$
in addition to the usual Higgs doublet $\Phi$ of the standard model,
motivated by the fact that this is the most economic,
model-independent way of generating neutrino masses through triplet
interactions. Then we have considered the most general scalar
potential of such a scenario, including a term $\azero\Phi\Phi
X^\dagger$. We show that, with just one triplet added, such a term
must be included if one has to avoid additional Goldstone bosons in
neutral sector. It is further demonstrated that $\azero$ must be real,
thus ruling out any additional source of CP-violation. We have also
obtained the field content requirement of a general model (with
doublets, complex-triplets, real-triplets) for having additional
sources of CP-violation in the tree-level potential.

We have gone on to examine the implications of the above trilinear
term in the mass matrices of the neutral scalars and pseudoscalars, as
also for the singly- and doubly-charged scalar masses in this
scenario. The $\azero$-dependent lower limits on the charged Higgs
mass are derived from bounds on $\rho$-parameter.

We find that, for small values of $|\delta|=|\azero|/\vtwo$ (where
$\vtwo$ is the triplet vev), the spectrum is of such nature as to
allow the decay of heavier scalars into lighter ones via gauge
interactions. For large $|\delta|$, on the other hand, the
doubly-charged, singly-charged and neutral pseudoscalar Higgses become
practically degenerate, while the two even-parity neutral scalars are
considerably lighter. This also emphasizes the possibility of the
decay of the singly-charged or neutral pseudoscalar states into the
neutral scalars.

The couplings of the various physical states in this scenario has been
studied in detail. It is found that, for small values of $|\delta|$,
the lightest neutral scalar field is dominated by triplet
contributions, and as such has extremely suppressed interactions with
fermion as well as gauge boson pairs.

{\bf{Acknowledgments}} 

AK acknowledges support from the research projects SR/S2/HEP-15/2003 of
DST, Govt.\ of India, and 2007/37/9/BRNS of DAE, Govt.\ of India. PD thanks 
the Department of Physics, University of Calcutta, for a visit. BM
thanks the Department of Physics, University of Calcutta, for a visit
under the UGC (UPE) scheme. The work of PD and BM has been partially supported
by the funds made available to the Regional Centre for Accelerator-based
Particle Physics (ReCAPP) by the Department of Atomic Energy, Govt.\ of
India.

\newpage
\vskip 20pt {\Large{\bf Appendix: Gauge and fermion couplings with
scalars}}
\label{secap}

Here we present a complete list of Feynman rules for couplings of the
physical scalars with gauge bosons and fermions. We define
$\cw=\cos\theta_{_W}$ etc.,
 and the mixing angles,
\begin{eqnarray}
\label{ap2}
\cplus &=& \frac{\vone}{\sqrt{\vone^2+4\vtwo^2}}, ~~~~ \splus =
\frac{2\vtwo}{\sqrt{\vone^2+4\vtwo^2}} \\ \cimg &=&
\frac{\vone}{\sqrt{\vone^2+8\vtwo^2}}, ~~~~ \simg =
\frac{2\sqrt2\vtwo}{\sqrt{\vone^2+8\vtwo^2}},
\end{eqnarray}
for the charged scalars and pseudo scalars with their respective gauge
states. For the real scalars we use the corresponding notations $\creal$ and
$\sreal$. In terms of the elements of ${\cal M}^{0R}$ in equation
(\ref{zzmatrixir}), and its eigenvalues $\Lambda_{\hlr}=2M^2_{\hlr}$ and
$\Lambda_{\hhr}=2M^2_{\hhr}$, the expression for $\creal$ and $\sreal$ can be
given as,
\begin{eqnarray}
\creal &=& {\cal M}^{0R}_{12}/\left(\sqrt{\left[{\cal M}^{0R}_{12}\right]^2 +
    \left[\Lambda_{\hlr} - {\cal M}^{0R}_{11}\right]^2}\right) \nonumber\\
    \sreal &=& {\cal M}^{0R}_{12}/\left(\sqrt{\left[{\cal
    M}^{0R}_{12}\right]^2 + \left[\Lambda_{\hhr} - {\cal
    M}^{0R}_{11}\right]^2}\right),
\label{realexps}
\end{eqnarray}
which we estimate numerically. While deducing the Feynman rules, we assume all
momenta (expressed in general as $\pmu$) to be incoming.

Following the above notations we now list the rules for the gauge-scalar
three-vertices;
\begin{eqnarray}
\label{apthree}
Z\hpp\hmm \futki i\left(g\frac{\cos2\theta_{_W}}{\cw}\right)
\left[\pmu(\hmm)-\pmu(\hpp)\right] \\ 
A\hpp\hmm \futki i\left(2e\right)
\left[\pmu(\hmm)-\pmu(\hpp)\right] \\ 
W^-W^-\hpp \futki i\left(2g^2\vtwo\right) \gab \\
ZZ\hlr \futki i\left(\frac{g^2}{\cw^2}\right) \left[\frac{\vone\creal}{2} +
  2\sqrt{2}\vtwo\sreal\right]\gab \\ 
ZZ\hhr \futki i\left(\frac{g^2}{\cw^2}\right) \left[-\frac{\vone\sreal}{2} +
  2\sqrt{2}\vtwo\creal\right]\gab \\ 
W^+W^-\hlr \futki ig^2 \left[\frac{\vone\creal}{2} +
  2\sqrt{2}\vtwo\sreal\right]\gab \\ 
W^+W^-\hhr \futki ig^2 \left[-\frac{\vone\sreal}{2} +
  2\sqrt{2}\vtwo\creal\right]\gab \\ 
ZW^-\hlp \futki \left(\frac{g^2}{\cw}\left[\cos2\theta_{_W} - 1\right]
\frac{\vone\cplus}{4} - \left[eg'+ \frac{g^2}{\cw}\right]\vtwo\splus\right)
\gab \\   
ZW^-\hhp \futki \left(-\frac{g^2}{\cw}\left[\cos2\theta_{_W} - 1\right]
\frac{\vone\splus}{4} - \left[eg'+ \frac{g^2}{\cw}\right]\vtwo\cplus\right)
\gab \\ 
AW^-\hlp \futki eg\left(\frac{\vone\cplus}{4} + \vtwo\splus\right) \gab \\ 
AW^-\hhp \futki eg\left(-\frac{\vone\splus}{4} + \vtwo\cplus\right) \gab \\ 
Z\hlp\hlm \futki i\left(g\frac{\cos2\theta_{_W}}{2\cw}\cplus^2 +
g'\sw\splus^2\right) \left[\pmu(\hlm)-\pmu(\hlp)\right] \\ 
Z\hlp\hhm \futki i\left(-g\frac{\cos2\theta_{_W}}{2\cw} +
g'\sw\right)\cplus\splus \left[\pmu(\hlm)-\pmu(\hlp)\right] \\  
Z\hhp\hhm \futki i\left(g\frac{\cos2\theta_{_W}}{2\cw}\splus^2 +
g'\sw\cplus^2\right) \left[\pmu(\hlm)-\pmu(\hlp)\right] \\  
A\hlp\hlm \futki ie \left[\pmu(\hlm)-\pmu(\hlp)\right] \\ 
A\hhp\hhm \futki ie \left[\pmu(\hhm)-\pmu(\hhp)\right] \\
Z\hlr\hli \futki \frac{g}{\cw}\left(\creal\cimg/2+\sreal\simg\right) 
\left[\pmu(\hlr)-\pmu(\hli)\right] \\
Z\hlr\hhi \futki \frac{g}{\cw}\left(\sreal\cimg/2-\creal\simg\right) 
\left[\pmu(\hlr)-\pmu(\hhi)\right] \\
Z\hhr\hli \futki \frac{g}{\cw}\left(-\creal\simg/2+\sreal\cimg\right) 
\left[\pmu(\hhr)-\pmu(\hli)\right] \\
Z\hhr\hhi \futki \frac{g}{\cw}\left(-\sreal\simg/2-\creal\cimg\right) 
\left[\pmu(\hhr)-\pmu(\hhi)\right] \\
W^-\hlm\hpp \futki i\left(g\splus\right) 
\left[\pmu(\hlm)-\pmu(\hpp)\right] \\
W^-\hhm\hpp \futki i\left(g\cplus\right) 
\left[\pmu(\hhm)-\pmu(\hpp)\right] \\
W^-\hlp\hlr \futki ig\left(\cplus\creal/2 + \splus\sreal/\sqrt2\right)
\left[\pmu(\hlr)-\pmu(\hlp)\right] \\
W^-\hhp\hlr \futki ig\left(-\splus\creal/2 + \cplus\sreal/\sqrt2\right)
\left[\pmu(\hlr)-\pmu(\hhp)\right] \\
W^-\hlp\hhr \futki ig\left(-\cplus\sreal/2 + \splus\creal/\sqrt2\right)
\left[\pmu(\hhr)-\pmu(\hlp)\right] \\
W^-\hhp\hhr \futki ig\left(\splus\sreal/2 + \cplus\creal/\sqrt2\right)
\left[\pmu(\hhr)-\pmu(\hhp)\right] \\
W^-\hlp\hli \futki g\left(\cplus\cimg/2 + \splus\simg/\sqrt2\right)
\left[\pmu(\hli)-\pmu(\hlp)\right] \\
W^-\hhp\hli \futki g\left(\splus\cimg/2 - \cplus\simg/\sqrt2\right)
\left[\pmu(\hli)-\pmu(\hhp)\right] \\
W^-\hlp\hhi \futki g\left(-\cplus\simg/2 + \splus\cimg/\sqrt2\right)
\left[\pmu(\hhi)-\pmu(\hlp)\right] \\
W^-\hhp\hhi \futki g\left(-\splus\simg/2 - \cplus\cimg/\sqrt2\right)
\left[\pmu(\hhi)-\pmu(\hhp)\right] 
\end{eqnarray}
The Feynman rules for gauge-scalar four-vertices are,
\begin{eqnarray}
\label{apfour}
ZZ\hpp\hmm \futki i\left(2g^2\frac{\cos^22\theta_{_W}}{\cw^2}\right)\gab \\ 
AZ\hpp\hmm \futki i\left(4eg\frac{\cos2\theta_{_W}}{\cw}\right)\gab \\ 
AA\hpp\hmm \futki i\left(8e^2\right)\gab \\
W^+W^-\hpp\hmm \futki ig^2\gab \\ 
ZZ\hlp\hlm \futki i\left(g^2\frac{\cos^22\theta_{_W}}{2\cw^2}\cplus^2 +
2g'^2\sw^2\splus^2\right)\gab \\ 
ZZ\hlp\hhm \futki -i\left(g^2\frac{\cos^22\theta_{_W}}{\cw^2} -
2g'^2\sw^2\right)\cplus\splus\gab \\ 
ZZ\hhp\hhm \futki i\left(g^2\frac{\cos^22\theta_{_W}}{2\cw^2}\splus^2 +
2g'^2\sw^2\cplus^2\right)\gab \\ 
AA\hlp\hlm \futki i\left(2e^2\right)\gab \\  
AA\hhp\hhm \futki i\left(2e^2\right)\gab \\  
AZ\hlp\hlm \futki i\left(eg\frac{\cos2\theta_{_W}}{\cw}\cplus^2 -
2eg'\sw\splus^2\right)\gab \\ 
AZ\hlp\hhm \futki -i\left(eg\frac{\cos2\theta_{_W}}{\cw} -
2eg'\sw\right)\cplus\splus\gab \\ 
AZ\hhp\hhm \futki i\left(eg\frac{\cos2\theta_{_W}}{\cw}\splus^2 -
2eg'\sw\cplus^2\right)\gab \\ 
W^+W^-\hlp\hlm \futki i\left(g^2\cplus^2/2 + 2g^2\splus^2\right)\gab \\ 
W^+W^-\hlp\hhm \futki i\left(-\frac{3}{2}g^2\cplus\splus\right)\gab \\ 
W^+W^-\hhp\hhm \futki i\left(g^2\splus^2/2 + 2g^2\cplus^2\right)\gab \\  
ZZ\hlr\hlr \futki i\left(g^2\frac{\creal^2}{2\cw^2} 
+ 2g^2\frac{\sreal^2}{\cw^2}\right)\gab \\
ZZ\hlr\hhr \futki i\left(g^2\frac{3\creal\sreal}{2\cw^2}\right)\gab \\
ZZ\hhr\hhr \futki i\left(g^2\frac{\sreal^2}{2\cw^2} 
+ 2g^2\frac{\creal^2}{\cw^2}\right)\gab \\
W^+W^-\hlr\hlr \futki i\left(g^2\creal^2/2 + 2g^2\sreal^2\right)\gab \\
W^+W^-\hlr\hhr \futki i\left(g^2\creal\sreal/2\right)\gab \\
W^+W^-\hhr\hhr \futki i\left(g^2\sreal^2/2 + 2g^2\creal^2\right)\gab \\ 
ZZ\hli\hli \futki i\left(g^2\frac{\cimg^2}{2\cw^2} 
+ 2g^2\frac{\simg^2}{\cw^2}\right)\gab \\
ZZ\hli\hhi \futki -i\left(g^2\frac{3\cimg\simg}{2\cw^2}\right)\gab \\
ZZ\hhi\hhi \futki i\left(g^2\frac{\simg^2}{2\cw^2} 
+ 2g^2\frac{\cimg^2}{\cw^2}\right)\gab \\
W^+W^-\hli\hli \futki i\left(g^2\cimg^2/2 + 2g^2\simg^2\right)\gab \\
W^+W^-\hli\hhi \futki -i\left(g^2\cimg\simg/2\right)\gab \\
W^+W^-\hhi\hhi \futki i\left(g^2\simg^2/2 + 2g^2\cimg^2\right)\gab \\ 
ZW^-\hpp\hlm \futki i\left(g^2\frac{\cos2\theta_{_W}}{\cw} 
- gg'\sw\right)\simg \gab \\ 
ZW^-\hpp\hhm \futki i\left(g^2\frac{\cos2\theta_{_W}}{\cw} 
- gg'\sw\right)\cimg \gab \\ 
AW^-\hpp\hlm \futki i\left(3eg\right)\simg \gab \\
AW^-\hpp\hhm \futki i\left(3eg\right)\cimg \gab \\
ZW^-\hlp\hlr \futki i\left(\frac{g^2}{\cw}\left[\cos2\theta_{_W} -
  1\right]\frac{\cplus\creal}{4} - \left[eg' + \frac{g^2}{\cw}\right]
\frac{\splus\sreal}{\sqrt{2}}\right) \gab \\ 
ZW^-\hlp\hhr \futki i\left(-\frac{g^2}{\cw}\left[\cos2\theta_{_W} -
  1\right]\frac{\cplus\sreal}{4} - \left[eg' + \frac{g^2}{\cw}\right]
\frac{\splus\creal}{\sqrt{2}}\right) \gab \\ 
ZW^-\hhp\hlr \futki i\left(-\frac{g^2}{\cw}\left[\cos2\theta_{_W} -
  1\right]\frac{\splus\creal}{4} - \left[eg' + \frac{g^2}{\cw}\right]
\frac{\cplus\sreal}{\sqrt{2}}\right) \gab \\
ZW^-\hhp\hhr \futki i\left(\frac{g^2}{\cw}\left[\cos2\theta_{_W} -
  1\right]\frac{\splus\sreal}{4} - \left[eg' + \frac{g^2}{\cw}\right]
\frac{\cplus\creal}{\sqrt{2}}\right) \gab \\
AW^-\hlp\hlr \futki i\left(eg\left[\frac{\cplus\creal}{2} +
  \frac{\splus\sreal}{\sqrt{2}}\right]\right) \gab \\ 
AW^-\hlp\hhr \futki i\left(eg\left[-\frac{\cplus\sreal}{2} +
  \frac{\splus\creal}{\sqrt{2}}\right]\right) \gab \\ 
AW^-\hhp\hlr \futki i\left(eg\left[-\frac{\splus\creal}{2} +
  \frac{\cplus\sreal}{\sqrt{2}}\right]\right) \gab \\ 
AW^-\hhp\hhr \futki i\left(eg\left[\frac{\splus\sreal}{2} +
  \frac{\splus\creal}{\sqrt{2}}\right]\right) \gab \\ 
ZW^-\hlp\hli \futki \left(\frac{g^2}{\cw}\left[\cos2\theta_{_W} -
  1\right]\frac{\cplus\cimg}{4} - \left[eg' + \frac{g^2}{\cw}\right]
\frac{\splus\simg}{\sqrt{2}}\right) \gab \\ 
ZW^-\hlp\hhi \futki \left(\frac{g^2}{\cw}\left[\cos2\theta_{_W} -
  1\right]\frac{\cplus\simg}{4} + \left[eg' + \frac{g^2}{\cw}\right]
\frac{\splus\cimg}{\sqrt{2}}\right) \gab \\ 
ZW^-\hhp\hli \futki \left(-\frac{g^2}{\cw}\left[\cos2\theta_{_W} -
  1\right]\frac{\splus\cimg}{4} - \left[eg' + \frac{g^2}{\cw}\right]
\frac{\cplus\simg}{\sqrt{2}}\right) \gab \\ 
ZW^-\hhp\hhi \futki \left(-\frac{g^2}{\cw}\left[\cos2\theta_{_W} -
  1\right]\frac{\splus\simg}{4} + \left[eg' + \frac{g^2}{\cw}\right]
\frac{\cplus\cimg}{\sqrt{2}}\right) \gab \\ 
AW^-\hlp\hli \futki eg\left(\frac{\cplus\cimg}{2} +
\frac{\splus\simg}{\sqrt{2}}\right) \gab \\ 
AW^-\hlp\hhi \futki eg\left(\frac{\cplus\simg}{2} -
\frac{\splus\simg}{\sqrt{2}}\right) \gab \\ 
AW^-\hhp\hli \futki eg\left(-\frac{\splus\cimg}{2} +
\frac{\cplus\simg}{\sqrt{2}}\right) \gab \\ 
AW^-\hhp\hhi \futki eg\left(-\frac{\splus\simg}{2} -
\frac{\cplus\cimg}{\sqrt{2}}\right) \gab \\ 
W^+W^+\hmm\hlr \futki i\left(\sqrt2g^2\sreal\right) \gab \\
W^+W^+\hmm\hhr \futki i\left(\sqrt2g^2\creal\right) \gab \\
W^+W^+\hmm\hli \futki -\left(\sqrt2g^2\simg\right) \gab \\
W^+W^+\hmm\hhi \futki \left(\sqrt2g^2\cimg\right) \gab.
\end{eqnarray}

Feynman rules for the fermion-scalar vertices are,
\begin{eqnarray}
\label{apferm}
\hlp \bar u_i d_j \futki \frac{ig}{\sqrt2M_W}
\left[m_{u_i}P_L - m_{d_j}P_R\right] V_{ij} \\
\hlm \bar d_j u_i \futki \frac{ig}{\sqrt2M_W}
\left[m_{u_i}P_R - m_{d_j}P_L\right] V^*_{ij} \\
\hhp \bar u_i d_j \futki \frac{ig\splus}{\sqrt2M_W\cplus}
\left[m_{u_i}P_L - m_{d_j}P_R\right] V_{ij} \\
\hhm \bar d_j u_i \futki \frac{ig\splus}{\sqrt2M_W\cplus}
\left[m_{u_i}P_R - m_{d_j}P_L\right] V^*_{ij} \\
\hli \bar u_i u_i \futki -\frac{g\cimg}{2M_W\cplus} m_{u_i} \gamma_5 \\
\hli \bar d_i d_i \futki \frac{g\cimg}{2M_W\cplus} m_{d_i} \gamma_5 \\
\hhi \bar u_i u_i \futki -\frac{g\simg}{2M_W\cplus} m_{u_i} \gamma_5 \\
\hhi \bar d_i d_i \futki \frac{g\simg}{2M_W\cplus} m_{d_i} \gamma_5 \\
\hlr \bar u_i u_i \futki \frac{ig\creal}{2M_W\cplus} m_{u_i} \\
\hlr \bar d_i d_i \futki \frac{ig\creal}{2M_W\cplus} m_{d_i} \\
\hhr \bar u_i u_i \futki \frac{ig\sreal}{2M_W\cplus} m_{u_i} \\
\hhr \bar d_i d_i \futki \frac{ig\sreal}{2M_W\cplus} m_{d_i} 
\end{eqnarray}

\end{document}